\documentclass[12pt]{article}
\usepackage{epsf,amssymb,psfrag,amsmath}
\catcode`\@=11
\def \be  {\begin{equation}}
\def \ee  {\end{equation}}
\def \ba  {\begin{eqnarray}}
\def \ea  {\end{eqnarray}}
\def \baa {\begin{eqnarray*}}
\def \eaa {\end{eqnarray*}}
\def \bb  {\begin {thebibliography} }
\def \eb  {\end{thebibliography}}
\def \lab #1 {\label{#1}}

\newcommand\re[1]{({\ref{#1}})}
\def \qqquad {\qquad\quad}
\def \qqqquad {\qquad\qquad}
\def \matrix #1 {\left(\begin{array}{cc} #1 \end{array}\right)}

\def \tr {\mathop{\rm tr}\nolimits}
\def \Im {\mathop{\rm Im}\nolimits}

\def \e  {\mathop{\rm e}\nolimits}
\newcommand\lr[1]{{\left({#1}\right)}}

\newcommand{\as}{\ifmmode\alpha_{\rm s}\else{$\alpha_{\rm s}$}\fi}
\newcommand{\asbar}{\ifmmode\bar{\alpha}_{\rm s}\else{$\bar{\alpha}_{\rm s}$}\fi}
\newcommand{\ft}[2]{{\textstyle\frac{#1}{#2}}}

\newcommand{\insertfig}[2]{\mbox{\epsfxsize=#1cm \epsfbox{#2.eps}}}

\font\cmss=cmss12 
\def\inbar{\,\vrule height1.5ex width.4pt depth0pt}
\def\IC{\relax\hbox{$\inbar\kern-.3em{\rm C}$}}
\def\IZ{\relax{\hbox{\cmss Z\kern-.4em Z}}}
\def\IR{{\hbox{{\rm I}\kern-.2em\hbox{\rm R}}}}

\def\IP{{\hbox{{\rm I}\kern-.2em\hbox{\rm P}}}}
\def\II{\hbox{{1}\kern-.25em\hbox{l}}}

\def\numberbysection{\@addtoreset{equation}{section}
                     \def\theequation{\thesection.\arabic{equation}}}
\numberbysection

\newbox\lett\newdimen\lheight\newdimen\lwidth
\def\ontop#1#2{\setbox\lett=\hbox{#2}\lheight\ht\lett
\multiply\lheight by 12 \divide\lheight by 10\relax%
\lwidth\wd\lett \multiply\lwidth by 8 \divide\lwidth by 10\relax #2\kern-\lwidth%
\raise\lheight\hbox{{$\scriptstyle #1$}}\kern.1ex}

\begin{document}

\begin{titlepage}
\begin{flushright}
\begin{tabular}{l}
LPT--Orsay--06--04 \\ 
ITEP-TH-02/06 \\
hep-th/0601112
\end{tabular}
\end{flushright}

\vskip2cm

\centerline{\large \bf Logarithmic scaling in gauge/string correspondence}

\vspace{1cm}

\centerline{\sc A.V. Belitsky$^a$, A.S. Gorsky$^b$, G.P. Korchemsky$^c$}

\vspace{10mm}

\centerline{\it $^a$Department of Physics and Astronomy, Arizona State
University} \centerline{\it Tempe, AZ 85287-1504, USA}

\vspace{3mm}

\centerline{\it $^b$Institute of Theoretical and Experimental Physics }
\centerline{\it  B. Cheremushkinskaya ul. 25, 117259 Moscow, Russia}

\vspace{3mm}

\centerline{\it $^c$Laboratoire de Physique Th\'eorique\footnote{Unit\'e
                    Mixte de Recherche du CNRS (UMR 8627).},
                    Universit\'e de Paris XI}
\centerline{\it 91405 Orsay C\'edex, France}

\def\thefootnote{\fnsymbol{footnote}}%
\vspace{1cm}

\centerline{\bf Abstract}

\vspace{5mm}

We study anomalous dimensions of (super)conformal Wilson operators at weak and strong
coupling making use of the integrability symmetry on both sides of the gauge/string
correspondence and elucidate the origin of their single-logarithmic behavior for long
operators/strings in the limit of large Lorentz spin. On the gauge theory side, we
apply the method of the Baxter $Q-$operator to identify different scaling regimes in
the anomalous dimensions in integrable sectors of (supersymmetric) Yang-Mills theory
to one-loop order and determine the values of the Lorentz spin at which the logarithmic
scaling sets in. We demonstrate that the conventional semiclassical approach based on
the analysis of the distribution of Bethe roots breaks down in this domain. We work out
an asymptotic expression for the anomalous dimensions which is valid throughout the
entire region of variation of the Lorentz spin. On the string theory side, the logarithmic
scaling occurs when two most distant points of the folded spinning string approach the
boundary of the AdS space. In terms of the spectral curve for the classical string sigma
model, the same configuration is described by an elliptic curve with two branching points
approaching values determined by the square root of the 't Hooft coupling constant. As a
result, the anomalous dimensions cease to obey the BMN scaling and scale logarithmically
with the Lorentz spin.

\end{titlepage}

\setcounter{footnote} 0

\newpage

\pagestyle{plain} \setcounter{page} 1

\section{Introduction}
\label{Introduction}

It is well known that in four-dimensional gauge theories the anomalous dimensions
of composite Wilson operators carrying a large Lorentz spin scale (at most)
logarithmically with the spin. This result is just one of the facets of a more
general Sudakov phenomenon \cite{Col89} and it can be traced back to the
existence of massless particles of spin one in the
spectrum -- the gauge fields. The logarithmic scaling of anomalous dimensions is
a universal feature of all gauge theories ranging from QCD to the maximally
supersymmetric $\mathcal{N} = 4$ Yang-Mills (SYM) theory. In particular, in the
simplest case of twist-two Wilson operators with large Lorentz spin $N \gg 1$,
the anomalous dimension behaves as (in the adjoint representation of the
$SU(N_c)$ group)~\cite{Kor88}
\be\label{gamma=cusp}
\gamma (\lambda) = 2 \Gamma_{\rm cusp}(\lambda) \ln N + \mathcal{O} (N^0) \, ,
\ee
where $\lambda=g_{\rm \scriptscriptstyle YM}^2 N_c$ is the 't Hooft coupling
constant and $\Gamma_{\rm cusp}(\lambda)$ is the so-called cusp anomalous
dimension~\cite{Pol80}. $\Gamma_{\rm cusp}(\lambda)$ is not universal however and
depends on the theory under consideration. It has numerous applications in
phenomenology of strong interactions and its calculation both at weak and strong
coupling regimes is one of the long-standing problems in gauge theories. At
present, the cusp anomaly is known in perturbation theory to the lowest three
orders~\cite{MocVerVog04,BerDixSmi05} and there exists a prediction at strong
coupling in the $\mathcal{N}=4$ SYM theory based on the gauge/string
correspondence~\cite{GubKlePol03,Kru02,BelGorKor03}.

The gauge/string correspondence \cite{Mal97} offers a powerful tool to study the
dynamics of four-dimen\-sional Yang-Mills theories at strong coupling. It
establishes a correspondence between Wilson operators in $\mathcal{N}=4$ SYM
theory and certain string excitations on the AdS${}_5\times$S${}^5$ background
\cite{BerMalNas02,GubKlePol03}. For operators carrying large quantum numbers (Lorentz spin,
isotopic $R-$charge,...) their scaling dimension at strong coupling can be found
as an energy of dual (semi)classical string configurations propagating on the
curved space. As was shown in Ref.~\cite{GubKlePol03}, the operators of twist two
with large Lorentz spin $N$ are dual to a folded string rotating with the angular
momentum $N$ on the AdS${}_3$ part of the target space. The resulting expression
for the twist-two anomalous dimension takes the form \re{gamma=cusp} with the
cusp anomalous dimension at strong coupling given by
\be\label{cusp}
\Gamma_{\rm cusp}(\lambda) \stackrel{\lambda
\gg 1}{=} \frac{\sqrt{\lambda} }{2\pi} + \mathcal{O}((\sqrt{\lambda})^0)\,.
\ee
Later, the dual string picture was generalized to Wilson operators of higher
twist in the $\mathcal{N}=4$ SYM theory carrying both large Lorentz spin and the
$R-$charge~\cite{FroTse03}. For operators built from holomorphic scalar fields
carrying a unit isotopic charge, the total $R-$charge equals the twist, $L$. For
such operators, in the dual picture the center-of-mass of the string rotates with
the angular momentum $L$ along a large circle of S${}^5$.

The important difference between the operators of twist two, $L=2$, and of higher
twist, $L\ge 3$, is that the latter are not uniquely specified by the total
Lorentz spin $N$. More precisely, for $L\ge 3$ there exist several
(superconformal) operators with the same $N$. These operators mix under
renormalization and the size of the mixing matrix rapidly increases with $L$ and
$N$. As a consequence, the anomalous dimensions of Wilson operators of high twist
$L\ge 3$ also depend on the integers $\ell = 1,2, \ldots$ which enumerate
eigenvalues of the mixing matrix (=anomalous dimensions) for a given Lorentz spin
$N$. For fixed $L$ and large $N$, possible values of the anomalous dimension
occupy a band~\cite{BelGorKor03}. The lower boundary of the band scales for
$N\to\infty$ as in \re{gamma=cusp} while the upper boundary scales as $\sim
L\,\Gamma_{\rm cusp}(\lambda) \ln N$. This implies that for operators of twist
$L=3,4,\ldots$ the minimal anomalous dimension has the same leading asymptotic
behavior for $N\to\infty$ as the twist-two anomalous dimension, Eq.~\re{gamma=cusp}.
This result is rather general and it holds true in a generic Yang-Mills theory
including QCD and $\mathcal{N}=4$ SYM theory.

In the present paper, we study the properties of the minimal anomalous dimensions
on both sides of the gauge/string correspondence in the limit of large twist
$L$ and Lorentz spin $N$. On the string side, the corresponding single-trace
Wilson operators are dual to a folded string spinning with large angular momentum
$N$ in the AdS${}_3$ part of the anti-de Sitter space and boosted along a large
circle on the sphere with a large angular momentum $L$ \cite{FroTse03}. The
energy $E$ of this classical string configuration defines the leading asymptotics
of the anomalous dimension  $\gamma(\lambda) = E - L - N$ of the dual Wilson
operator in the $\mathcal{N}=4$ SYM theory in the strong coupling regime in
planar approximation. The string theory provides a definite prediction for
$\gamma(\lambda)$ as a function of $L$ and $N$. One finds that $\gamma(\lambda)$
takes different forms in three regimes~\cite{GubKlePol03,FroTse03}:
\begin{itemize}
\item For $N \ll L$, in the ``short'' string limit
\be
\label{1st} \gamma (\lambda) = \lambda \frac{m^2}{2} \frac{N}{L^2} + \ldots
\ee
\item For $N \gg L$, in the ``long'' string limit
\be\label{3rd}
\gamma (\lambda) =\left \{
\begin{array}{l}
{\displaystyle \frac{\lambda}{2\pi^2} \frac{m^2}{L} \ln^2({N}/{L}) + \ldots \,
,\qquad
\mbox{for $\xi_{\rm str} < 1$}\,,} \\[3mm]
{\displaystyle \frac{\sqrt{\lambda}}{\pi}m\ln ({N}/\sqrt{\lambda}) +
\ldots \, ,\qquad \mbox{for $\xi_{\rm str} \gg 1$}\, ,}
\end{array}
\right.
\ee
\end{itemize}
with the parameter $\xi_{\rm str}$ defined as $\xi_{\rm str} =
\lambda\ln^2(N/L)/L^2$. Here the integer $m$ counts the number of times the
string is folded onto itself. The minimal anomalous dimension corresponds to a
single-folded string, $m=1$. In that case, for $\xi_{\rm str} \gg 1$, the leading
asymptotic behavior of $\gamma(\lambda)$ does not depend on the twist $L$ and is
the same as for $L = 2$ operators, Eqs.~\re{gamma=cusp} and \re{cusp}. For
$\xi_{\rm str} < 1$, the role of the twist $L$ is to create the ``BMN window'',
i.e., a region in the parameter space in which the anomalous dimension has an
expansion in powers of the BMN coupling $\lambda^\prime \equiv \lambda/(\pi
L)^2$~\cite{BerMalNas02}. It is believed that the first few terms in the
expansion of $\gamma(\lambda)$ in powers of $\lambda'$ should match similar
expressions for the anomalous dimensions of Wilson operators of twist $L$ and
spin $N$ obtained in the $\mathcal{N}=4$ SYM theory in the weak coupling
regime~\cite{Tse03}.

On the gauge theory side, the calculation of anomalous dimensions of higher twist
operators with large Lorentz spin turns out to be an extremely nontrivial task in
a generic Yang-Mills theory even to one-loop order due to a large size of the
mixing matrix. The problem can be overcome thanks to hidden integrability
symmetry of the dilatation operator~\cite{BraDerMan98,Bel99,DerKorMan99}, which
maps the one-loop mixing matrix for Wilson operators of twist $L$ belonging to
the so-called holomorphic $SL(2)$ sector into a Hamiltonian of the Heisenberg
magnet of length $L$ and spin $s$ determined by the conformal spin of the quantum
fields (for a review see Ref.~\cite{Bel04}). This observation allows one to
calculate the exact eigenspectrum of anomalous dimensions of Wilson operators of
arbitrary twist and Lorentz spin in integrable sectors of Yang-Mills theory by
means of the Quantum Inverse Scattering Method~\cite{TakFad79}. In the
$\mathcal{N}=4$ SYM theory, the minimal anomalous dimension of Wilson operators
built from $L$ scalar fields and carrying the Lorentz spin $N$ can be identified
to one-loop accuracy as the minimal energy in the eigenspectrum of the $SL(2)$
Heisenberg magnet of length $L$ and the total spin $N+Ls$ with $s=\ft12$
\cite{BeiFroStaTse03}. The gauge/string correspondence suggests that the minimal
anomalous dimension defined in this way should match the relations \re{1st} and
\re{3rd} in the thermodynamic limit $L\to \infty$.

It follows from \re{1st} and \re{3rd} that the anomalous dimensions of higher
twist operator depend on a ``hidden'' parameter $\xi_{\rm str} = \lambda
\ln^2(N/L)/L^2$ and their behavior at strong coupling is different for $\xi_{\rm
str}< 1$ and $\xi_{\rm str} \gg 1$. For $\xi_{\rm str}\gg 1$ the anomalous
dimension does not have a perturbative expansion in the BMN coupling
$\lambda^\prime$ and scales as $\sim \ln N$. On the gauge theory side, previous
studies~\cite{BeiFroStaTse03} of the Bethe ansatz for the $SL(2)$ spin chain in
the thermodynamic limit $L \gg 1$ led to the expression for $\gamma(\lambda)$
which coincides with \re{1st} in the limit of short strings and with the first
relation in \re{3rd} in the limit of long strings. They did not reveal however
neither any trace of the second, logarithmic regime in \re{3rd}, nor appearance
of a new parameter similar to $\xi_{\rm str}$. This fact is in contradiction with
our expectations that the minimal anomalous dimension of higher twist operators
should scale logarithmically to all loops as $N\to\infty$, Eq.~\re{gamma=cusp}.
The goal of the present study is to unravel the logarithmic scaling of the
anomalous dimension both in the gauge and string theory and to understand the
physical meaning of the parameter $\xi_{\rm str}$ and its counter-part $\xi$ on
the gauge theory side.

We shall revisit the calculation of the energy of the $SL(2)$ Heisenberg magnet
of spin $s=\ft12,1,\ft32$ using the method of the Baxter $Q-$operator
\cite{Bax72} as a main tool and demonstrate that the one-loop anomalous dimension
in integrable sectors of Yang-Mills theory has the following scaling behavior in
the thermodynamic limit $L\to\infty$:
\begin{itemize}
\item For $N \ll L$
\be
\label{1st-g} \gamma (\lambda) = \lambda \frac{m^2}{4s} \frac{N}{L^2} + \ldots \,
,
\ee
\item For $N \gg L$
\be\label{3rd-g}
\gamma (\lambda) =\left \{
\begin{array}{l}
{\displaystyle \frac{\lambda}{2\pi^2} \frac{m^2}{L} \ln^2({N}/{L}) + \ldots \,
,\qquad
\mbox{for $\xi < 1$}\,,} \\[3mm]
{\displaystyle \frac{{\lambda}}{2\pi^2}\, m\ln N +
\ldots \, ,\qqqquad~ \mbox{for $\xi \gg 1$}\, ,}
\end{array}
\right.
\ee
\end{itemize}
depending on the parameter $\xi = \ln(N/L)/L$. The minimal anomalous dimension
corresponds to $m=1$. Here $s$ equals the conformal spin of the field entering
the Wilson operator, i.e., $s=\ft12, 1, \ft32$ for scalar, gaugino and gluon
fields, respectively. For scalar operators, Eq.~\re{1st-g} and the first relation
in \re{3rd-g} coincide with similar expressions in \re{1st} and \re{3rd},
respectively. Notice that the anomalous dimension in \re{3rd-g} does not depend
on the spin $s$ for $N\gg L$ which suggests that the two regimes in \re{3rd-g}
are universal in all gauge theories. This is indeed the case for $\xi \gg 1$ since
the coefficient in front of $2\ln N$ coincides with the cusp anomalous dimension
at weak coupling, Eq.~\re{gamma=cusp}.

For $N\gg L$ and $\xi < 1$, the one-loop anomalous dimension exhibits a novel
double logarithmic behavior \re{3rd-g}. It was first discovered from the string
theory considerations~\cite{FroTse03} and was later reproduced on the gauge
theory side~\cite{BeiFroStaTse03}. A natural question arises whether similar
contributions arise at higher loops and whether they can be resummed to all
loops. The gauge/string correspondence suggests that in the $\mathcal{N}=4$ SYM
theory the anomalous dimension in the region $N \gg L \gg 1$ admits a BMN-like
expansion (with $\lambda'=\lambda/(\pi L)^2$ and $\xi = \ln(N/L)/L$)
\be\label{gen-exp}
\gamma (\lambda) = L \sum_{n=1}^\infty\lr{ \lambda^\prime \ln^2\frac{N}{L} }^n
c_n(\xi)+ \ldots\,,
\ee
where the coefficient functions $c_n(\xi)$ do not depend on the coupling constant
and have the following asymptotics for $\xi\to 0$ and $\xi\to\infty$
\be\label{coeff-fun}
c_n(\xi)=c_{0,n}+c_{1,n} \xi + \mathcal{O}(\xi^2) \,,\qquad c_n(\xi) =\mathcal{O}
(1/\xi^{2n-1} )
\ee
with $c_{0,n}=(-1)^n(-\ft12)_n/n!$. To one-loop order, Eq.~\re{gen-exp} matches
(for $m=1$) both relations in \re{3rd-g}. In addition, for $\xi\to \infty$ the
coefficient in front of $\lambda^n$ in the right-hand side of \re{gen-exp} scales
as $\sim \ln N$ and determines the $n-$loop correction to the cusp anomalous
dimension in the weak coupling regime, Eq.~\re{gamma=cusp}. In the strong
coupling regime, upon the substitution $\xi=\xi_{\rm str}^{1/2}/\sqrt{\lambda}$,
the perturbative series in \re{gen-exp} can be resummed to all loops into the
following expression (for $\lambda\to\infty$ and $\xi_{\rm str}=\lambda\ln^2(N/L)/L^2
=\rm fixed$)
\be
\gamma (\lambda) =  L \sum_{n=1}^\infty \lr{ \lambda^\prime \ln^2\frac{N}{L} }^n
c_n(0)+ \ldots = L \left[ \sqrt{1 + \lambda^\prime \ln^2\frac{N}{L}} - 1 \right] +
\ldots\,,
\ee
where the ellipsis stands for subleading corrections. For $N\to\infty$ this
relation also reproduces the leading asymptotic behavior of the anomalous
dimension in the last regime in \re{3rd} and, as a consequence, it leads to the
expression for the cusp anomalous dimension at strong coupling, Eq.~\re{cusp}.

As was already mentioned, the one-loop anomalous dimension \re{gen-exp} coincides
with the energy of the $SL(2)$ spin chain of length $L$ in the thermodynamic
limit $L\to\infty$. The latter can be found within the Bethe Ansatz approach by
systematically expanding the energy in powers of $1/L$ with a help of known
semiclassical methods \cite{Sut95}. To leading order of the semiclassical
expansion, the Bethe roots condense on two symmetric intervals on the real axis
$[-a,-b]\cup [b,a]$, with the boundaries $a$ and $b$ being functions of $L/N$.
This leads~\cite{BeiFroStaTse03} to the expression for the energy given in \re{1st}
for $N \ll L$ and in the first relation in \re{3rd} for $N\gg L$. It is believed
that subleading corrections to the energy are suppressed by powers of $1/L$ and,
therefore, are small. We demonstrate that this assumption is only justified for $\xi =
\ln(N/L)/L < 1$, while for $\xi\gg 1$ the semiclassical expansion of the energy
becomes divergent indicating the change of asymptotic behavior of the anomalous
dimension, Eq.~\re{3rd}. The reason why the semiclassical expansion fails is that
the two cuts $[-a,-b]$ and $[b,a]$ collide at the origin, that is $b\to 0$ for
$\xi\to\infty$, and the Bethe roots have a nonvanishing distribution at the
origin. As a consequence, for $N\gg L$ the semiclassical corrections to the
anomalous dimension run in powers of $\xi$. To one-loop order they are described
in \re{gen-exp} by the function $c_1(\xi)$. We argue that the semiclassical
series for $c_1(\xi)$ is divergent for $\xi>1$ and propose an approach which
circumvents this difficulty and allows one to determine this function for
arbitrary $\xi$. The resulting expression for the one-loop anomalous dimension is
valid in the thermodynamic limit throughout the entire interval of $N$ and
reproduces correct logarithmic behavior for $N \gg L$, Eq.~\re{3rd-g}.

On the string theory side, in the dual picture of the folded string spinning in
the AdS${}_3\times$S${}^1$ part of the target space, the logarithmic behavior of
the anomalous dimension at strong coupling is associated with the classical
string configuration which has two spikes approaching the boundary of the AdS
space. Thanks to classical integrability of the string equations of motion
\cite{ManSurWad02}, the same configuration is described by the spectral
(elliptic) symmetric curve endowed with a meromorphic differential of
quasimomentum possessing a double pole at $x=\pm \sqrt{\lambda'}$ and having a
prescribed asymptotic behavior at the origin and infinity~\cite{ZakMik78,Kri94,KazZar04}.
The branching points of the curve, $\pm b_{\rm str}$ and $\pm a_{\rm str}$, depend
on the ratio $L/N$ and the coupling constant $\lambda'$. We show that for $N \gg L$
and $\xi_{\rm str} < 1$ the branching points admit a regular expansion in powers
of $\lambda'$ and, as a consequence, the anomalous dimension exhibits the BMN scaling,
Eq.~\re{3rd}. For $\xi_{\rm str} \gg 1$, $b_{\rm str}$ approaches its minimal value
$\sqrt{\lambda'}$ so that the inner boundaries of two cuts $[-a_{\rm str},-b_{\rm
str}]$ and $[b_{\rm str},a_{\rm str}]$ coincide with the position of poles of the
momentum differential and cannot collide. This nonanalyticity manifests itself
through the appearance of $\sqrt{\lambda}$ prefactor in the logarithmic behavior
of the anomalous dimension at strong coupling, Eq.~\re{3rd}.

Our consequent presentation is organized as follows. In Sect.~\ref{SectAnDim}, we
outline a general framework for analysis of one-loop anomalous dimensions in the
thermodynamic limit. It is based on the semiclassical expansion of solutions to
the Baxter equation~\cite{PasGau92,Kor95,Smi98}. In Sect.~\ref{ThermodynamicSection},
we apply the semiclassical approach to determine the minimal anomalous dimension
in the thermodynamic limit and demonstrate that the semiclassical expansion breaks
down for $\ln(N/L) \gg L $ due to collision of cuts. Then, we present an approach
to go consistently beyond the semiclassical expansion and use it to describe the
minimal anomalous dimension for large Lorentz spin. In Sect.~\ref{ConclusionSection},
we analyze the asymptotic behavior of the anomalous dimensions at strong coupling
based on the string sigma model consideration. Section~\ref{StringSigmaModel}
contains concluding remarks. Some technical details of our calculations are
summarized in the Appendix.

\section{Anomalous dimensions in gauge theory}
\label{SectAnDim}

Let us start with the calculation of one-loop anomalous dimensions of
(super)conformal Wilson operators of arbitrary twist $L$ and Lorentz spin $N$
belonging to integrable sectors of (supersymmetric) Yang-Mills theories. For
quantum fields transforming in the adjoint representation of the gauge group, the
operators under considerations have the following generic form
\be\label{O-def}
\mathcal{O}_{N,L}(0) = \sum_{k_1+\ldots+k_L =N} c_{k_1\ldots k_L} \tr
\left\{D_+^{k_1} X(0)D_+^{k_2} X(0)\ldots D_+^{k_L} X(0)\right\},
\ee
where $X(0)$ stands for the so-called ``good'' component of quantum fields of a
definite helicity in the underlying gauge theory, $D_+=D_\mu n^\mu$ is the
covariant derivative projected onto the light-cone, $n_\mu^2=0$. The expansion
coefficients $c_{k_1\ldots k_L}$ are fixed from the condition for
$\mathcal{O}_{N,L}(0)$ to have an autonomous scale dependence, i.e., Eq.\
\re{O-def} has to be an eigenstate of the one-loop dilatation operator.
Integrability allows one to map the one-loop anomalous dimension of the operators
\re{O-def} into energy $\varepsilon$ of the noncompact $SL(2)$ Heisenberg spin
chain of length $L$ and the total spin $N + Ls$
\cite{BraDerMan98,Bel99,DerKorMan99,Bei04},
\be\label{gamma=energy}
\gamma(\lambda) = \frac{\lambda}{8\pi^2}\, \varepsilon +
\mathcal{O}(\lambda^2)\,.
\ee
Here the (half-)integer $s$ is given by the conformal spin \cite{Bel04} of the
quantum field $X(0)$, that is, $s=1/2$ for scalars, $s=1$ for gaugino fields of
helicity $\pm 1/2$ and $s=3/2$ for gauge fields of helicity $\pm 1$.

\subsection{Exact solution}

Let us first describe the exact solution for the energy $\varepsilon$ of the
$SL(2)$ magnet of length $L$ and single-particle spin $s$ in each site. We shall
employ the method of the Baxter $Q-$operator \cite{Bax72} which proves to be
convenient for analyzing various semiclassical limits of $\varepsilon
=\varepsilon(N,L)$ including the limit of the large spin $N$ and length
$L$~\cite{Kor95,Smi98}. The method relies on the existence of an operator $Q(u)$
which acts on the Hilbert space of the $SL(2)$ spin chain and is diagonalized by
all eigenstates of the magnet for arbitrary complex parameter $u$. Discussing the
energy spectrum it suffices to study the eigenvalues of the $Q-$operator that we
shall denote by $Q(u)$. The same function $Q(u)$ determines the wave function of
the magnet in the representation of Separated Variables \cite{Skl90}
\footnote{For an interpretation of the $Q-$operator in string theory see
Ref.~\cite{Gor03}.}. This allows one to analyze $Q(u)$ in the semiclassical limit
with the help of the WKB machinery well-known from quantum mechanics and, then,
determine $\varepsilon(N,L)$.

By construction, $Q(u)$ satisfies the second-order finite-difference equation
\be\label{Baxter-eq}
(u + is)^L Q (u + i) + (u - is)^L Q(u - i) = t_L(u) Q(u) \, ,
\ee
which can be thought of as a Schr\"odinger equation for a single-particle wave
function in the representation of Separated Variables~\cite{Skl90}. Here $t_L(u)$
is a polynomial in $u$ of degree $L$ with coefficients given by conserved charges
\be\label{t_L}
t_L(u) = 2 u^L + q_2 u^{L-2} + \ldots + q_L
\ee
The lowest integral of motion $q_2$ is related to the total spin of the $SL(2)$
chain, $N+Ls$,
\be\label{q2}
q_2 = -(N + Ls) (N + Ls - 1) + L s (s-1)
\ee
with $N=0,1,\ldots$.

In what follows we shall refer to Eq.\ \re{Baxter-eq} as the Baxter equation.
Taken alone, it does not fix the function $Q(u)$ and it has to be supplemented
by an additional condition that $Q(u)$ has to be polynomial in $u$ \cite{FadKor94}.
Examining the asymptotic behavior of both sides of \re{Baxter-eq} for $u\to\infty$,
it is easy to see that the degree of $Q(u)$ is fixed by the total spin $N$ and,
therefore, up to an overall normalization, one can write
\be\label{Q-polynom}
Q(u) = \prod_{k=1}^N (u-\lambda_k)\,.
\ee
One substitutes this ansatz into \re{Baxter-eq}, takes $u=\lambda_k$ and finds
that the roots $\lambda_1,\ldots,\lambda_N$ satisfy the Bethe equations
\be\label{Bethe-roots}
\lr{\frac{\lambda_k+is}{\lambda_k-is}}^L=\prod_{j=1,j\neq k}^N
\frac{\lambda_k-\lambda_j-i}{\lambda_k-\lambda_j+i}\,.
\ee
Solving the Baxter equation \re{Baxter-eq} supplemented by \re{Q-polynom} one
obtains quantized values of the charges $q_3,\ldots,q_L$ and evaluates the
corresponding energy and quasimomentum as \cite{FadKor94}
\be\label{Energy-Baxter}
\varepsilon = i\lr{\ln Q(is)}'-i\lr{\ln Q(-is)}'\,,\qquad \e^{i\theta} =
\frac{Q(is)}{Q(-is)}\,.
\ee
Replacing $Q(u)$ by its expression \re{Q-polynom} one verifies that these relations
coincide with those coming from the Algebraic Bethe Ansatz~\cite{TakFad79}
\be\label{Energy-ABA}
\varepsilon = \sum_{k=1}^N \frac{2s}{\lambda_k^2+s^2}
\, ,
\qquad \e^{i\theta} = \prod_{k=1}^N \frac{\lambda_k-is}{\lambda_k+is}
\, .
\ee
The cyclic symmetry of the single-trace operators \re{O-def} imposes an
additional selection rule for the eigenstates of the spin magnet,
$\e^{i\theta}=1$. Equations \re{Energy-Baxter} and \re{Energy-ABA} allow one to
calculate the energy of the spin chain and, then, obtain the one-loop anomalous
dimension of Wilson operators \re{O-def} with a help of \re{gamma=energy}.

\subsection{Quasiclassical approach}

Let us examine the Baxter equation \re{Baxter-eq} for $N+Ls \gg 1$. In this
limit, the charge $q_2$ takes large negative values and one can apply
semiclassical techniques~\cite{Kor95} to construct the solution to
\re{Baxter-eq}. To go over to the semiclassical limit, we introduce two scaling
parameters
\be\label{beta}
\eta=(N+Ls)^{-1}\,,\qquad \beta =  s L \eta =\frac{s L}{ N+Ls } \,.
\ee
By definition, $0\le \beta \le 1$ with the boundary values corresponding to
\be\label{limits}
\beta\stackrel{L\ll N}{\longrightarrow} 0
\, , \qqquad
\beta\stackrel{L\gg N}{\longrightarrow} 1
\, .
\ee
The parameter $\eta \ll 1$ will play the role of the Planck constant. One
rescales the spectral parameter as $u=x/\eta$ and introduces the eikonal phase
(the Hamilton-Jacobi ``action'' function) $S(x)$ as
\be\label{Q-ansatz}
Q(x/\eta) = \eta^{-N}\,\exp\lr{\frac1{\eta} S(x)}  \,,\qquad S(x) = \eta
\sum_{k=1}^N \ln (x-\eta \lambda_k)\,.
\ee
The energy and the quasimomentum in Eq.~\re{Energy-Baxter} are then given in
terms of the function $S(x)$ by the following expressions
\be\label{E-general}
\varepsilon = i \left[S'(i\beta/L) - S'(-i\beta/L)\right] \,,\qquad \e^{i\theta}
= \exp\left\{\frac1{\eta}[S(i\beta/L) - S(-i\beta/L)]\right\} \,.
\ee
It also proves convenient to introduce a notation for the ``effective potential''
\be\label{tau-expansion}
\tau(x) =(\eta/x)^{L}\,t_L(x/\eta) = 2 + \frac{\widehat q_2}{x^2} +
\frac{\widehat q_3}{x^3} + \ldots + \frac{\widehat q_L}{x^L}
\, ,
\ee
with $\widehat q_k=q_k \eta^k$.

In the semiclassical approach~\cite{PasGau92,Kor95} one assumes that the
function $S(x)$ and the integrals of motion $q_k$ (with $k=3,\ldots,L$) admit a
systematic expansion in powers of $\eta$
\be\label{WKB-ansatz}
S(x) = S_0(x) + \eta\, S_1(x) + \ldots\,,\qquad \widehat q_k = \widehat
q_k^{\scriptscriptstyle (0)} + \eta \, \widehat q_k^{\scriptscriptstyle (1)} +
\ldots\, .
\ee
It is tacitly assumed that the expansion of $S(x)$ is convergent and each term is
uniformly bounded.%
\footnote{As we will show in Sect.~\ref{ThermodynamicSection}, this assumption is
justified for $\ln(N/L) < L$ and is invalid otherwise.} This leads to the expansion
of the effective potential \re{tau-expansion}, $\tau(x)= \tau_0(x) + \eta \tau_1(x)
+ \ldots$ with
\be\label{tau0}
\tau_0(x) =  2 - \frac1{x^2}+\frac{\widehat q_3^{\scriptscriptstyle
(0)}}{x^3}+\ldots +\frac{\widehat q_L^{\scriptscriptstyle (0)}}{x^L} \,,\qquad
\tau_1(x) = \frac{\widehat q_3^{\scriptscriptstyle (1)}}{x^3}+\ldots
+\frac{\widehat{q}_L^{\scriptscriptstyle (1)}}{x^L}\,,\qquad \ldots
\ee
One substitutes \re{WKB-ansatz} into the Baxter equation \re{Baxter-eq} and
equates the coefficients in front of powers of $\eta$ to get to leading order
\be\label{p-def}
2 \cos p(x) = \tau_0 (x)\,,\qquad p(x) = S_0^\prime (x) + \frac{\beta}x\,.
\ee
In the finite-gap theory, the function $p(x)$ defines the Bloch-Floquet
multiplier in an auxiliary linear problem for the Baker-Akhiezer function and has
the meaning of the (quasi)momentum \cite{NovManPitZak84,FadTak87,BabBerTal03}.
For the first subleading term in the semiclassical expansion one finds in a
similar manner
\ba\label{S1-Baxter}
S_1^\prime (x) = - \frac{p^\prime (x)}{2} \cot p (x) - \frac{1}{2 \sin p(x)}
\left( \tau_1(x) + \frac{\beta (1 - s)}{2 x^2} \tau_0(x) \right) \, .
\ea
It is straightforward to derive the subleading terms $S_{k\ge 2}'(x)$ but we will
not need them for our purposes. The obtained expressions for the action functions
$S'_0(x)$ and $S'_1(x)$ depend on yet unknown conserved charges $\widehat
q_k^{\scriptscriptstyle (0)}$ and $\widehat q_k^{\scriptscriptstyle (1)}$,
respectively. Quantization conditions for these charges follow from the requirement
for $Q(x/\eta)$, Eq.~\re{Q-ansatz}, to be a single valued function of $x$.

According to \re{tau0}, $\tau_0(x)$ is a polynomial of degree $L$ in $1/x$.
Solving \re{p-def}, one finds that the momentum $p(x)$ is, in general, a
double-valued function on the complex $x-$plane with the square-root branching
points $x_j$ obeying the condition $\tau_0 (x_j)=\pm 2$. It is convenient to
introduce the function $y(x)=2 \sin p(x)$ and define a complex curve~\cite{Kor95}
\be\label{curve}
\Gamma_L: \qquad y^2 = 4-\tau_0^2(x)\,,\qquad \tau_0(x) = 2 - \frac1{x^2} +
\frac{\widehat q_3^{\scriptscriptstyle (0)}}{x^3} + \ldots + \frac{\widehat
q_L^{\scriptscriptstyle (0)}}{x^L}\,.
\ee
For arbitrary complex $x$, except the branching points $y(x_j)=0$, the relation
\re{curve} defines two values for $y(x)$. Then, $y(x)$ being a double-valued
function on the complex $x-$plane, becomes a single-valued function on the
hyperelliptic Riemann surface defined by the complex curve $\Gamma_L$. This
surface has a genus $L-2$ and is realized by gluing together two copies of the
complex $x-$plane along the cuts running between the branching points $x_{2j-1}$
and $x_{2j}$.

For the $SL(2)$ magnet the Bethe roots verifying Eq.\ \re{Bethe-roots} take real
values only, ${\rm Im} \lambda_k =0$. In the semiclassical limit, $\eta \to 0$,
they condense on finite intervals on the real axis where the momentum $p(x)$
takes purely imaginary values~\cite{Kor95}. In terms of the hyperelliptic curve
\re{curve}, this corresponds to $y^2 \le 0$, or
\be\label{cuts}
\tau_0^2(x) \ge 4\,,\qquad \mbox{for $x\in \mathcal{S}=[x_{2L-2},x_{2L-3}]\cup
\ldots \cup [x_4,x_3] \cup [x_2,x_1]  $}\,,
\ee
where $x_1>x_2> \ldots >x_{2L-2}$ and one of the intervals contains the origin.
The total number of intervals in \re{cuts} equals $L-1$ and the end points $x_j$
are just the branching points of the complex curve \re{curve}, $\tau_0^2(x_j)=4$.
As follows from \re{curve}, the curve can be parameterized by the set of $2L-2$
real branching points as
\be\label{y2-roots}
y^2 = \frac4{x^2} \prod_{j=1}^{2L-2} \lr{1-\frac{x_j}{x}}\,.
\ee
The intervals $\mathcal{S}$ have the meaning of regions where the classical
motion of the system occurs in the separated variables\footnote{In the
finite-gap theory~\cite{NovManPitZak84}, the same intervals have the meaning of
forbidden zones in the auxiliary linear problem for the Baker-Akhiezer function.}.
Later on we shall encounter the situation when, say, $j^{\rm th}$ interval shrinks
into a point, $x_{2j}=x_{2j-1}$, so that the motion on this interval is frozen at
the classical level. In what follows we shall refer to $x_{2j}=x_{2j-1}$ as a
double point.

\begin{figure}[t]
\begin{center}
\mbox{
\begin{picture}(0,135)(230,0)
\put(0,0){\insertfig{16}{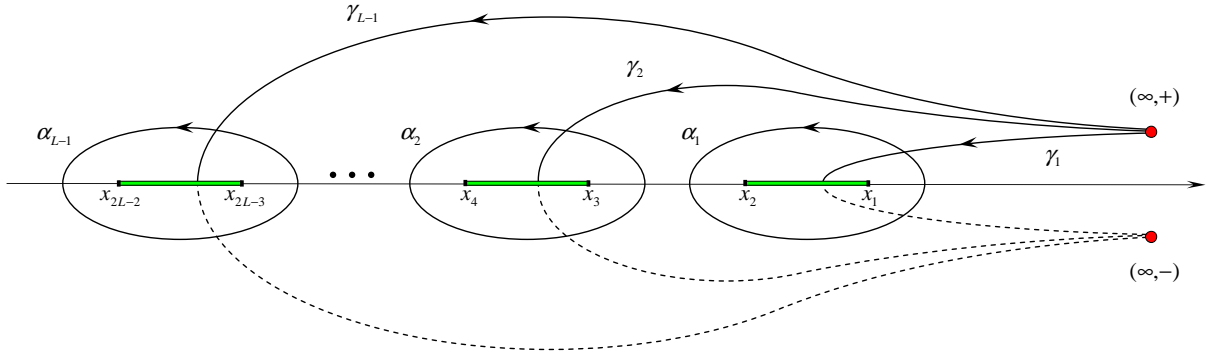}}
\end{picture}
}
\end{center}
\caption{ \label{CyclesFig} The definition of the $\alpha-$cycles and
$\gamma-$contours on the Riemann surface $\Gamma_L$. The dashed lines represent
the part of the path on the lower sheet of the surface. }
\end{figure}

The leading term of the semiclassical expansion, $S_0(x)$, is determined by the
momentum $p(x)$, Eq.~\re{p-def}. As follows from its definition \re{p-def}
\be\label{p-exp}
p(x) = 2i \ln \frac{\sqrt{\tau_0(x)+2}-\sqrt{\tau_0(x)-2}}{2}\,.
\ee
The momentum $p(x)$ takes purely imaginary values on the intervals \re{cuts} and
its values at the end point of the $j^{\rm th}$ interval coincide,
$p(x_{2j-1})=p(x_{2j})$, with $\e^{i p(x_{2j})}=\pm 1$ for $\tau_0(x_{2j})=\pm
2$, respectively. Continuing $p(x)$ to the complex $x-$plane one finds that
$p'(x)$ is an analytical function on the complex plane with cuts running along
the intervals \re{cuts}. It defines a meromorphic differential on $\Gamma_L$
\be\label{dp}
dp = p'(x)\, dx = -\frac{\tau_0'(x)}{\sqrt{4-\tau_0^2(x)}}dx\,.
\ee
From \re{curve} and \re{p-exp} one finds $\tau_0(x) = 2- 1/x^2 +
\mathcal{O}(1/x^3)$ so that $\e^{p(\infty)}=1$ and
\be\label{p-infinity}
dp \sim\mp \frac{dx}{x^2}\,,\qquad \mbox{for $x\to\infty$}
\ee
where `$-/+$' correspond to the upper/lower sheet of $\Gamma_L$. According to the
definition \re{p-def}, $p(\infty)$ is defined modulo $2\pi$. Choosing the
normalization condition $p(\infty) = 0$, one finds that at the end points of the
$j^{\rm th}$ interval in \re{cuts}, the momentum takes the values $p(x_{2j})
=p(x_{2j-1}) = - \pi j$ for $x_{2j}>0$. As a consequence, the differential $dp$
satisfies the normalization conditions~\cite{SmiRes83,NovManPitZak84}
\be\label{p-periods}
2 \int^{x_{2j-1}}_{x_{2j}} dx\, p'(x) = - \oint_{\alpha_j} dp = 0\,,\qquad 2
\int_{x_{2j-1}}^{\infty} dx\, p'(x) = - \int_{\gamma_j} dp = -2\pi j\,.
\ee
Here in both equations, the integration in the left-hand side goes over the upper
sheet of $\Gamma_L$. In the right-hand side of the first relation, the
differential $dp$ is integrated over the $\alpha_j-$cycle encircling the interval
$[x_{2j},x_{2j-1}]$ in the anticlockwise direction. The contour $\gamma_j$ in the
second relation starts on the upper sheet above $x = \infty$ crosses the same
interval and then goes to infinity on the lower sheet (see Fig.\
\ref{CyclesFig}).

The obtained expressions for $S_0'(x)$ and $S_1'(x)$, Eqs.~\re{p-def} and
\re{S1-Baxter}, respectively, depend on the conserved charges $\widehat q_k$, yet
to be determined. To work out the quantization conditions for $\widehat q_k$ one
examines the first derivative of the eikonal phase \re{Q-ansatz}
\be\label{S-prime}
S'(x) = \eta \sum_{k=1}^N \frac1{x-\eta \lambda_k}\,.
\ee
The discontinuity of $S'(x)$ across the cuts \re{cuts} gives the distribution
density of rescaled Bethe roots $\eta \lambda_k$. Assuming that the roots $\eta
\lambda_k$ take finite values for $\eta\to 0$, one finds the behavior of $S'(x)$
at infinity on the upper, physical sheet of $\Gamma_L$ as
\be\label{S-res-inf}
S'(x) \sim \frac{\eta N}{x}=\frac{1-\beta}{x}\,,\qquad {\rm for}\ x \to \infty\,,
\ee
with $\eta$ and $\beta$ defined in \re{beta}. Replacing $S(x)$ by its
semiclassical expansion \re{WKB-ansatz} and matching the
coefficients in front of powers of $\eta$ one obtains%
\footnote{Later on we shall consider solutions to the Baxter equation satisfying
$Q(u)=Q(-u)$, or equivalently $S(x)=S(-x)$. For such solutions, $S_k'(x)$ are odd
functions of $x$ and their asymptotics at infinity involves odd powers of $x$
only, that is, $S_{k \ge 1}'(x) \sim 1/{x^3}$.}
\be\label{as-infinity}
S_0'(x) \sim \frac{1-\beta}{x}\,,\qquad S_{1}'(x) \sim \frac1{x^2} \,,\qquad
\ldots
\ee
%
%
According to \re{Q-polynom}, the total number of Bethe roots equals $N$. For a
given energy level they are distributed on $L-1$ intervals \re{cuts}. Denoting by
$\ell_j$ the number of Bethe roots on the $j^{\rm th}$ interval, one has
\be\label{a-period}
\frac1{2\pi i}\oint_{\alpha_j} dx\, S'(x) = \eta \ell_j=
\frac{\ell_j}{N+Ls}\,,\qquad (j=1,\ldots,L-1)
\, .
\ee
The sum of all $\alpha-$cycles is homologous to zero and, as a consequence, the sum
of all $\alpha-$periods is given by the residue of $S'(x)$ at infinity. Together
with \re{S-res-inf} this leads to $N=\ell_1 + \ldots+\ell_{L-1}$.

Solving the quantization conditions \re{a-period} one can determine quantized
values of the conserved charges $\widehat q_k$. According to \re{a-period} they
depend both on the scaling parameter $\eta$ and the set of nonnegative integers
$\ell_1,\ldots,\ell_{L-1}$. Replacing $\widehat q_k$ in \re{tau0}, \re{p-def} and
\re{S1-Baxter} by their quantized values, one constructs semiclassical expression
for $S(x)$ and, then, applies \re{E-general} to determine the energy and
quasimomentum,
\be
\widehat q_k = \widehat q_k(\ell_1,\ldots,\ell_{L-1};\eta) \, , \qquad
\varepsilon = \varepsilon (\ell_1,\ldots,\ell_{L-1};\eta)\,.
\ee
Explicit form of these relations for $L=3$ can be found in Refs.\
\cite{Kor95,BraDerMan98,Bel99,DerKorMan99}. In particular, the quantized values
of the energy and conserved charges exhibit remarkable regularity and form
trajectories. The flow parameter along each trajectory is given by the total
spin $\eta=(N+Ls)^{-1}$ while the integers $\ell_1,\ldots,\ell_{L-1}$ enumerate
the trajectories and encode a nontrivial analytic structure in the eigenspectrum.
For given $N$, the total number of trajectories equals the number of partitions
of $N$ into the sum of $L-1$ nonnegative integers $\ell_1,\ldots,\ell_{L-1}$.
This is in a perfect agreement with the number of irreducible components entering
the tensor product of $L$ copies of the $SL(2)$ modules~\cite{BelGorKor03}.

\subsection{Minimal energy trajectory}
\label{CollisionCuts}

In the semiclassical approach described in the previous section, the energy of
the spin magnet, or equivalently the one-loop anomalous dimension of Wilson
operators, is parameterized by the complex curve $\Gamma_L$, Eq.~\re{curve}. The
genus of the curve $g=L-2$ is defined by the length of the spin chain whereas its
moduli are determined by the quantized values of the conserved charges $\widehat
q_k$ which depend in their turn on the set of integers $\ell_1,\ldots,\ell_{L-1}$.
Going over to different parts of the spectrum amounts to specifying the integers
$\ell_1,\ldots,\ell_{L-1}$.

In this paper we are interested in the eigenstates possessing the minimal
possible energy for a given total spin $N$. Such eigenstates belong to a
particular trajectory to which we shall refer as the minimal energy trajectory
(see Fig.\ \ref{Fig-roots-energy10} below). To describe it one has first to
identify the corresponding integers $\ell_1,\ldots,\ell_{L-1}$. We remind that
the energy \re{Energy-ABA} is determined by zeros of the function $Q(u)$, or
equivalently the Bethe roots. In the thermodynamical limit they condense on the
intervals \re{cuts}. For a given total spin $N$, the minimal energy is realized
when the Bethe roots are located on two symmetric cuts most distant from the
origin \cite{Sut95}, that is, $\ell_1 = \ell_{L-1}=N/2$ and $\ell_j=0$ for
$j=2,\ldots,L-2$. The corresponding $Q(u)$ should be an even function,
$Q(u)=Q(-u)$. According to \re{Energy-Baxter}, \re{Q-polynom} and \re{Baxter-eq},
such states automatically satisfy the cyclic symmetry condition $\e^{i\theta}=1$
and possess the following quantum numbers
\be\label{q-odd}
N = {\rm even}\,,\qquad q_{2k+1} = 0
\, ,
\ee
with $0 \le k \le (L-1)/2$. The fact that $\ell_j=0$ implies that $j^{\rm th}$
interval does not contain Bethe roots and, therefore, it shrinks into a point,
$x_{2j-1}=x_{2j}$. From the point of view of separated variables, this means
that classically all but two collective degrees of freedom are frozen and the
classical motion is confined to the two intervals with $\ell_1 = \ell_{L-1} =N/2$.
For the complex curve \re{y2-roots}, this implies that all branching points
except four, $x_1=-x_{2L-2}$ and $x_2=-x_{2L-3}$, become the double points,
$x_{2j-1}=x_{2j}$, and the curve $\Gamma_L$ reduces to the elliptic curve (see
Eq.~\re{curve-reduced} below).

Let us apply the semiclassical approach to obtain the expression for the energy
for $L\gg 1$ along the minimal energy trajectory as a function of $N$. To begin
with, we consider the ground state of the $SL(2)$ spin chain. It has the total
spin $N=0$ and is described by a trivial solution to the Baxter equation
\re{Baxter-eq}, $Q(u)=1$, or equivalently $S(x)=0$. From \re{Energy-Baxter}, the
corresponding energy is $\varepsilon = 0$ and the integrals of motion can be read
off from \re{Baxter-eq} upon substituting $Q(u)=1$. To leading order in $\eta=
1/(sL)$ the transfer matrix and momentum, Eq.~\re{p-def}, look like
\be
\label{N=0-roots}
\tau_0(x)=2 \cos(1/x)\,,\qquad p(x) = 1/x \,.
\ee
Matching these expressions into \re{curve} and \re{y2-roots} one finds that for
$L\to \infty$ all branching points of the spectral curve \re{y2-roots} are double
points, $x_{2j-1} = x_{2j} = 1/(\pi j)$.

Let us now consider the minimal energy eigenstates with $N \gg 1$  and $L \gg 1$
and distinguish two limiting cases: (i) $N/L={\rm fixed}$ and (ii) $N/L \gg 1$,
corresponding to $0 < \beta < 1$ and $\beta\to 0$, respectively, Eq.~\re{beta}.
We recall that the function $Q(u)$ has exactly $N$ roots, Eq.~\re{Q-polynom}.
Assuming that the Bethe roots scale as $\lambda_k \sim 1/\eta$, one finds from
\re{Energy-ABA} that for $N/L={\rm fixed}$ the energy should behave as
$\varepsilon \sim N/L^2\sim 1/L$. Indeed, one can obtain the same scaling of
$\varepsilon$ by naively expanding the energy \re{E-general} in powers of $1/L$
\be\label{E-naive-expansion}
\varepsilon = -\frac{2\beta}{L} S''(0) + \mathcal{O}(\beta^3/L^3) =
-\frac{2\beta}{L} \left[ S_0''(0) + \eta S_1''(0)+ \mathcal{O}(\eta^2)\right]
\, .
\ee
Here in the second relation we replaced $S(x)$ by its semiclassical expansion
\re{WKB-ansatz}. In a similar manner, one obtains for the quasimomentum
\re{E-general}
\be
\theta = 2s \left[S_0'(0) + \eta S_1'(0) + \mathcal{O}(\eta^2)\right]=0\,,
\ee
where we took into account that $S_k(x)$ are even functions of $x$ for the
minimal energy states
\be\label{as-zero}
S_0'(0) = 0 \,,\qquad S_1'(0) = 0\,, \quad \ldots
\, .
\ee
Then, one finds from \re{p-def} the asymptotic behavior of the momentum around
the origin on the upper (``physical'') sheet of the Riemann surface \re{curve} as
\be\label{p-zero}
p(x) = \frac{\beta}{x} +  \mathcal{O}(x)\,,
\ee
so that the differential $dp$ has a double pole above $x=0$. It is important to
stress that the relations \re{E-naive-expansion} and \re{p-zero} were obtained
under the assumption that $S'(x)$ is regular at the origin.
For the minimal energy eigenstates, all but two intervals in \re{cuts} shrink
into the double points and the Bethe roots condense on two symmetric cuts on the
real axis that we shall denote as $[-a,-b]$ and $[b,a]$. Then, the spectral curve
\re{y2-roots} reduces to
\be\label{curve-reduced}
y^2 = \tilde y^2 \left[ \frac2{x^3}\prod_j \lr{1-\frac{x^2_{2j}}{x^2}} \right]^2
\,,\qquad \tilde y^2 = (x^2-a^2)(x^2-b^2) \,,
\ee
where we took into account that for the minimal energy states the branching
points appear in pairs of opposite sign. The parameters $a$, $b$ and $x_{2j}$ are
determined by the condition for $S_0'(x)$, Eq.~\re{p-def}, to be an analytical
function on the complex $x-$plane with two symmetric cuts on the real axis and
prescribed asymptotic behavior at $x=0$ and $x\to\infty$, Eqs.~\re{as-zero} and
\re{as-infinity}, respectively \cite{SmiRes83,NovManPitZak84}. Equation
\re{E-naive-expansion} implies that the energy has the BMN scaling in the
thermodynamic limit, $\varepsilon \sim 1/L$. This relation is in an apparent
contradiction with the well-known fact that for $N \gg L$ the anomalous dimension
should scale logarithmically $\varepsilon \sim \ln N$ with the prefactor being
$L$ independent. To understand the reason for this discrepancy, one notices that
the relation \re{E-naive-expansion} is valid provided that $S'(x)$ is analytical
in the vicinity of $x=0$, or equivalently, there is no accumulation of the Bethe
roots around the origin. For $N,\, L \to \infty$ with $N/L={\rm fixed}$ this is
indeed the case but, as we will argue in the next section, the situation
drastically changes for $N\gg L$.

As a hint, let us consider the eigenstates with $L={\rm fixed}$, $N\to\infty$ and
the charges satisfying \re{q-odd}. In this case, the transfer matrix $\tau_0(x)$,
Eq.~\re{tau0}, is an even polynomial of degree $L$ which scales for $x\to 0$ as
\be\label{tau-x=0}
\tau_0(x) \sim \frac{\widehat q_L^{\scriptscriptstyle (0)}}{x^L}
\ee
provided that $\widehat q_L^{\scriptscriptstyle (0)}\neq 0$. If $\widehat
q_L^{\scriptscriptstyle (0)}$ vanishes, the asymptotics of $\tau_0(x)$ is
governed by the first nonvanishing charge $\widehat q_{2k}^{\scriptscriptstyle
(0)} \neq 0$ with $2k < L$. Substituting \re{tau-x=0} into \re{dp} one finds the
leading asymptotic behavior of the momentum for $x\to 0$ on the upper sheet of
\re{curve} as
\be
dp \sim  i L \frac{dx}{x}\,,
\ee
yielding $p(x) \sim i L \ln x$. According to \re{beta}, $\beta\sim sL/N \to 0$
for $N\to\infty$ and one finds from \re{p-def} the asymptotics of $S^\prime_0(x)$
above $x=0$ on the physical sheet of \re{curve} as
\be\label{S0-log}
S_0'(x) \sim i L \ln x\,.
\ee
Comparing this relation with \re{as-zero}, one concludes that $S_0'(x)$ is no
longer analytical at the origin due to accumulation of rescaled Bethe roots at
$x=0$. Finally, one substitutes \re{S0-log} into \re{E-general} and obtains the
energy
\be\label{E-ln}
\varepsilon \sim 2 L \ln(L/\beta) \sim 2 L \ln N\,.
\ee
Notice that the coefficient in front of $\ln N$ in the right-hand side of
\re{E-ln} is determined by the leading asymptotic behavior of the transfer matrix
\re{tau-x=0} for $x\to 0$. For $\widehat q_L^{\rm (0)}\neq 0$ this coefficient
takes the maximal possible value. For $\widehat q_L^{\scriptscriptstyle (0)}
=\ldots=\widehat q_{2m+2}^{\scriptscriptstyle (0)} =0$, the transfer matrix
\re{tau0} scales as $\tau_0(x)\sim \widehat q_{2m}^{\scriptscriptstyle (0)}
/x^{2m}$ (recall that the charges with odd indices vanish, Eq.~\re{q-odd})
leading to
\be
\label{E-m} \varepsilon \sim 4m \ln N \,,\qquad \widehat
q_{2m+2}^{\scriptscriptstyle (0)}=\ldots=\widehat q_{L}^{\scriptscriptstyle (0)}
= 0\,.
\ee
The values of the remaining charges $\widehat q_{4}^{\scriptscriptstyle
(0)},\ldots,\widehat q_{2m}^{\scriptscriptstyle (0)}$ will be determined below
(see Eq.~\re{q-expression}). Throughout this paper we are interested in the
eigenstates with the minimal energy for given $N$. Obviously, they correspond to
$m=1$, that is,
\be\label{Emin-naive}
\varepsilon^{(m=1)} \sim 4 \ln N \, , \qquad \widehat q_4^{ (0)} = \ldots =
\widehat q_L^{\scriptscriptstyle (0)}=0 \, .
\ee
Here the superscript `$\scriptstyle (0)$' refers to the leading order
approximation of the semiclassical expansion and $\widehat
q_{2k}^{\scriptscriptstyle (0)}=0$ does not necessarily imply that $\widehat
q_{2k}=0$ but rather $\widehat q_{2k}= \mathcal{O}(\eta)$, or equivalently
$q_{2k} \ll (-q_2)^{k}$.

The transfer matrix $\tau_0(x)$, Eq.~\re{tau0}, corresponding to the minimal
energy state \re{Emin-naive} is given by
\be\label{tau-min}
\tau^{(m=1)}_{0}(x)=2-\frac1{x^2}
\ee
and the spectral curve \re{curve} looks like
\be\label{Gamma-min}
\Gamma_L^{(m=1)}: \qquad y^2 = \lr{x^2-\frac1{4}}\frac4{x^4}\,.
\ee
It is easy to see that $\Gamma_L^{(m=1)}$ coincides with the complex curve for
the spin chain of length $2$, that is $\Gamma_{L=2}$, Eq.~\re{curve}. Indeed,
choosing $\widehat q_{2m+1}^{\scriptscriptstyle (0)}=\ldots=\widehat
q_{L}^{\scriptscriptstyle (0)} = 0$ in \re{curve} one effectively descends from
an infinitely long spin chain (for $L\to\infty$) to the one with finite length
$2m$. Comparing \re{Gamma-min} with \re{y2-roots} we conclude that for $m=1$ all
but two branching points condense at the origin, $b=x_{2j}=0$, and the two
remaining (resolved) branching points are located at $\pm 1/2$. For $m\ge 2$, one
can show~\cite{KorKri97} that the complex curve corresponding to \re{E-m} is
given by
\be\label{Gamma-m}
\Gamma_L^{(m\ge 2)}: \qquad  y^2=
\lr{x^2-\frac1{4m^2}}\frac4{x^4}\prod_{j=1}^{m-1}
\lr{1-\frac{x^2_{2j}}{x^2}}^2\,,
\ee
with $1/x_{2j}=2m\cos\lr{\frac{\pi j}{2m}}$. Comparison with \re{curve-reduced}
yields $a=1/(2m)$ and $b=0$, that is, the two cuts $[-a,-b]$ and $[b,a]$ collide
at the origin. Notice that $\Gamma_L^{(m\ge 2)}$ has exactly $2m$ double points
satisfying $x_{2j}^2 > 1/(2m)^2$ while the remaining double points in
\re{curve-reduced} belonging to the interval $[-b,b]$ condensed at $x=0$ as $b\to
0$.

The complex curves \re{Gamma-min} and \re{Gamma-m} have genus zero and, as a
consequence, the momentum $p'(x)$ is expressed in terms of elementary functions.
Indeed, one substitutes \re{Gamma-min} and \re{Gamma-m} into \re{dp} and obtains
the following expression for the momentum for arbitrary $m$
\be\label{dp-reduced}
dp = -\frac{dx}{x\sqrt{x^2-\frac1{(2m)^2}}}\,,\qquad
p (x) = \int^x_{\infty} dp =
2 m \arcsin \left( \frac{1}{2m x} \right) \, .
\ee
Together with \re{p-def} and \re{tau0} this leads to the following expressions
for the integrals of motion
\be\label{q-expression}
\widehat  q_{2n}^{\scriptscriptstyle (0)} = \frac{2 (- 1)^{n} m^{1 - 2 n} \Gamma
(m + n) }{ \Gamma (2 n + 1) \Gamma (m - n + 1)} \, .
\ee
We conclude from \re{dp-reduced} that for $m=1$, for the eigenstate with the
minimal energy, in the limit $N\to\infty$ and $L=\rm fixed$, the momentum $p'(x)$
is an analytical function on the complex plane with the square root cut
$[-\ft12,\ft12]$ and a {\sl simple} pole at the origin. This should be compared
with the analytical properties of the momentum in the region $N/L={\rm fixed}$ in
which case $p'(x)$ has a {\sl double} pole at $x=0$, Eq.~\re{p-zero}.

\begin{figure}[t]
\begin{center}
\mbox{
\begin{picture}(0,70)(240,0)
\put(0,0){\insertfig{17}{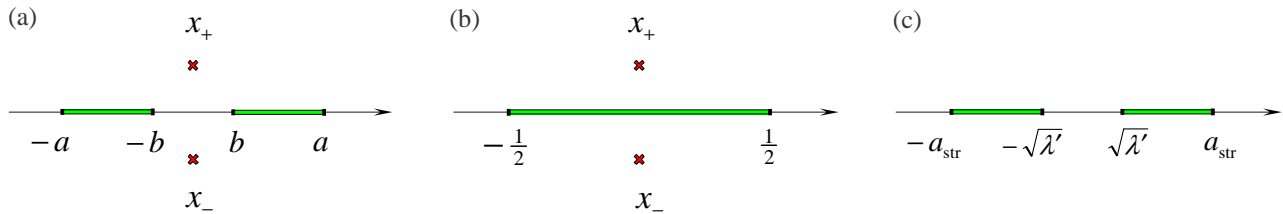}}
\end{picture}
}
\end{center}
\caption{\label{CollisionCutsFig}%
Symmetric two-cut configuration (a) resulting in the BMN scaling of the anomalous
dimension in gauge theory for $\xi < 1$. For $\xi\gg 1$ the two cuts collide at
the origin (b) yielding the logarithmic scaling. The same configuration in string
sigma model (c) for $\xi_{\rm str}\gg 1$ -- the minimal value for the inner end
of the cut is given by the BMN coupling $\sqrt{\lambda^\prime}$ which prevents
the cuts to collide.}
\end{figure}

We demonstrated in this section that the analytical properties of the momentum
$p(x)$ and the action function $S'(x)$ are quite different in the two limits
mentioned above. For $0 < \beta < 1$, the Bethe roots condense on two symmetric
intervals $[-a,-b]$ and $[b,a]$ with the end-points $a$ and $b$ depending on
$\beta$. For $\beta\to 0$, one has $b\to 0$ and $a\to \ft12$ so that the two cuts
collide and form a single cut $[-\ft12,\ft12]$ (see Fig.\ \ref{CollisionCutsFig}
(a) and (b)). This leads to different scaling behavior of the energy in these two
limits, Eqs.~\re{E-naive-expansion} and \re{Emin-naive}, respectively. The
question remains what happens in the intermediate region of the parameter
$\beta=Ls/(N+Ls)$, Eq.~\re{beta}, and how important the corrections $\sim \eta$
to the energy \re{E-naive-expansion} are. We shall demonstrate in the next
section that the semiclassical expansion of the energy \re{E-naive-expansion}
becomes divergent for $\beta\to 0$ due to the collision of cuts and work out an
asymptotic expression for the energy valid for small $\beta$.

\section{Symmetric two-cut solution in thermodynamic limit}
\label{ThermodynamicSection}

In this section, we construct the minimal energy trajectory in the thermodynamic
limit $L \gg 1$ for arbitrary values of the spin $N$. To begin with, we work out
the semiclassical expansion of solutions $Q(u)$ to the Baxter equation
\re{Baxter-eq} for $N/L={\rm fixed}$ and, then, extend the analysis to the region
$N \gg L$.

\subsection{Semiclassical expansion at leading order}
\label{WKBloSection}

To leading order of the semiclassical expansion \re{WKB-ansatz} and
\re{Q-ansatz}, the function $S_0'(x)$ is determined by the momentum $p(x)$,
Eq.~\re{p-def}. The latter is fixed by the condition that $dp(x)$ should be a
meromorphic differential on the complex curve \re{curve-reduced} with prescribed
asymptotic behavior at infinity, Eq.~\re{p-infinity}, and at the origin,
Eq.~\re{p-zero}, on the upper sheet of $\Gamma_L$
\be
\label{p-asym}
dp \stackrel{x\to\infty}{\sim} -\frac{dx}{x^2}\,,\qqqquad dp \stackrel{x\to
0}{\sim} dx \left[-\frac{\beta}{x^2} + \mathcal{O}(x^0)\right]\,,
\ee
with $\beta=Ls/(N+Ls)$, Eq.~\re{beta}. Being combined together, these conditions
lead to the following expression~\cite{KazZar04}
\be\label{dp-res}
dp = \frac{-1+\frac{\beta ab}{x^{2}}}{\sqrt{(x^2-a^2)(x^2-b^2)}}\,dx\,.
\ee
It defines $dp$ as a meromorphic differential on the elliptic curve $\tilde y^2 =
(x^2-a^2)(x^2-b^2)$, Eq.~\re{curve-reduced}, which has a pair of double poles
above $x=0$ on both sheets of the corresponding Riemann surface. The values of
the parameters $a$ and $b$ are fixed by the normalization conditions
\re{p-periods}
\be\label{cond}
\int_{b}^a dp = 0\,,\qqqquad \int_a^\infty dp = - \pi m\,.
\ee
Here the integer $m$ defines the position of the interval $[b,a]$ inside
\re{cuts}. We remind that in the limit we are interested in, the complex curve
$\Gamma_L$ gets reduced to the genus one curve \re{curve-reduced}. Namely, all
cuts in \re{cuts} except two, $[-a,-b]$ and $[b,a]$, shrink into points. The
integer $m-1$ counts how many collapsed cuts are situated to the right from the
interval $[b,a]$ on the real axis. The energy depends on $m$ and, as we will see
in a moment, it takes the minimal value for $m=1$, that is, when all shrunken
cuts are located inside the interval $[-b,b]$.

Solving \re{cond} one finds the positions of the cuts as a function of the parameter
$\beta$, Eq.~\re{beta}
\be\label{ab}
a=\frac1{2m\,\mathbb{E}(\tau)}\,,\qquad b =
\frac{\beta}{2m}\frac1{\mathbb{K}(\tau)} \,,\qquad
\beta=\sqrt{1-\tau}\frac{\mathbb{K}(\tau)}{\mathbb{E}(\tau)}\,,
\ee
where $\mathbb{K}(\tau)$ and $\mathbb{E}(\tau)$ are elliptic integrals of the
first and second kind, respectively, and the modular parameter is defined as (for
$b \le a$)
\be\label{tau}
\tau = 1 -\frac{b^2}{a^2}\,.
\ee
From \re{dp-res} and \re{p-def} one obtains the leading term in the semiclassical
expansion as
\be\label{SO-zero}
S_0''(x) = p'(x) + \frac{\beta}{x^2}
=
\frac{-1+\frac{\beta ab}{x^{2}}}{\sqrt{(x^2-a^2)(x^2-b^2)}} + \frac{\beta}{x^2}
\, ,
\ee
with the parameters $a$ and $b$ given by \re{ab}. One verifies that $S_0''(x)$ has
a regular expansion around $x=0$ which is in agreement with our expectations that
the Bethe roots condense on the intervals $[-a,-b]\cup [b,a]$ and there is no
accumulation of roots at the origin. The energy \re{E-naive-expansion} is determined
by the leading term in the expansion of this expression~\cite{BeiFroStaTse03}
\be\label{Energy-2-cut}
\varepsilon
=
-
\frac{2\beta}{L} \left[{\frac {1}{ab}} - \frac{\beta}2\lr{\frac1{a^2}+\frac1{b^2}} \right]
=
\frac{(2m)^2}{L}
\mathbb{K}(\tau) \left[ (2-\tau) \mathbb{K}(\tau) - 2 \mathbb{E}(\tau)\right]
\, .
\ee
The dependence of $\varepsilon$ on the total spin $N$ and conformal spin $s$
enters into this expression through the parametric  dependence of the modular
parameter $\tau$, Eq.~\re{ab}, on the scaling parameter $\beta=Ls/(N+Ls)$,
Eq.~\re{beta}.

\begin{figure}[t]
\begin{center}
\mbox{
\begin{picture}(0,225)(150,0)
\put(0,0){\insertfig{10}{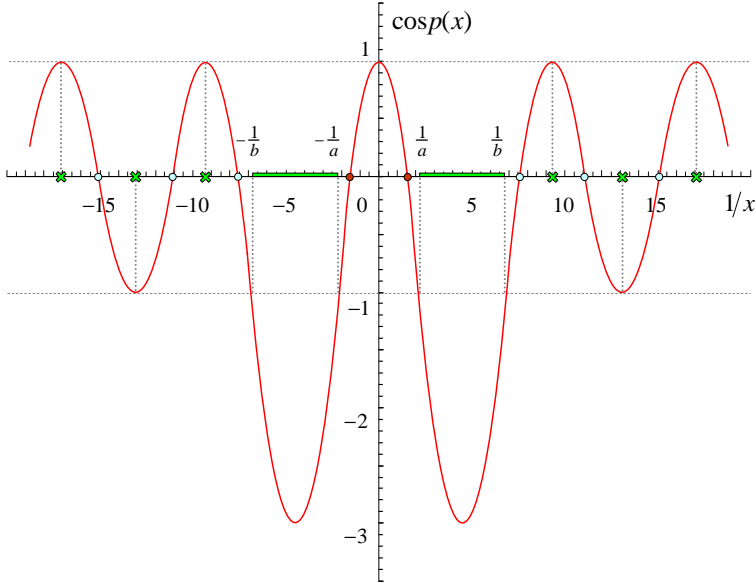}}
\end{picture}
}
\end{center}
\caption{\label{Fig-semi} The transfer matrix $\cos p(x)=\tau_0(x)/2$ as a
function of $1/x$ for the symmetric two-cut solution for $\beta=3/4$ and $m=1$.
The cuts $[-a,-b] \cup [b,a]$ with $a=0.45$ and $b=0.15$ correspond to $\cos^2
p(x)> 1$.  The double points $x_{2j}$ are denoted by crosses, $\cos p(x_{2j})=\pm
1$ and $1/x_{2j}^2 > 1/b^2$. The ``large'' and ``small'' roots of the transfer
matrix, $\cos p (\delta_n) = 0$, are shown by full and light blobs,
respectively.}
\end{figure}

We observe that the minimal energy corresponds to $m=1$, that is, when all double
points $x_{2j}$, Eq.~\re{curve-reduced}, are located inside the interval
$[-b,b]$. It is straightforward to find their position on the real axis with the
help of \re{N=0-roots} and \re{dp-res}. Since the total number of double points
is infinite for $L\to \infty$ and they occupy a compact interval, $x_{2j}$ should
be a smooth function of $j$. Differentiating the second relation in
\re{N=0-roots} with respect to $j$, one finds
\be\label{dx}
\frac{dx_{2j}}{dj} = \frac{\pi} {p'(x_{2j})} = - x_{2j}^2 \frac{\pi} {\beta}
\left[ 1+ \mathcal{O}(x_{2j}^2)\right]\,,
\ee
where we substituted the momentum by its value at the origin \re{p-asym}. It
follows from \re{dx} that for large $j$ the double points $1/x_{2j}$ are
equidistantly distributed on the real axis (see Fig.~\ref{Fig-semi})
\be\label{equid}
\frac{1}{x_{2j}} = \frac{\pi}{\beta} j + \mathcal{O}(1/j)\,.
\ee
We remind that the double points $x_{2j}$ verify the relation $\tau_0(x_{2j})=\pm
2$. It is instructive to compare \re{equid} with a similar relation $1/{x_{2j}} =
{\pi}j$ for the momentum \re{N=0-roots} corresponding to the state with $N=0$, or
equivalently $\beta=1$. Equation \re{equid} suggests that close to the origin and
away from the cuts, $x^2\ll b^2$, the transfer matrix for the two-cut solution
looks like $\tau_0(x) = 2 \cos p(x) \sim 2 \cos(\beta /x)$. Matching this
relation into the general expression for $\tau_0(x)$, Eq.~\re{tau-expansion}, we
find that the integrals of motion scale as
\be\label{q-scaling}
\widehat q_{2n} \sim 2\frac{(-1)^{n}}{(2n)!}\beta^{2n}\,,\qquad (n \gg 1) \, ,
\ee
where $\widehat q_{2n}=q_{2n}/(N+Ls)^{2n}$. For $\beta=1$, or equivalently $N=0$,
this relation is exact for all $\widehat q_{2k}$. For $\beta\to 0$, or
equivalently $N\to\infty$, it cannot be exact since, by definition, $\widehat q_2
= -1+\mathcal{O}(\eta)$. It instead indicates that higher charges take
anomalously small values $\widehat q_{2k}=\mathcal{O} (\eta^{2k})$ with $\eta\sim
1/N$. This is in agreement with our expectations that for $\beta=0$ the transfer
matrix reduces to \re{tau-min}.

Let us examine the expression for the energy \re{Energy-2-cut} in two limiting
cases $\beta\to 1$ and $\beta\to 0$, or equivalently $N \ll L$ and $N \gg L$,
respectively, Eq.~\re{limits}. According to \re{ab}, the corresponding values of
the modular parameter are $\tau\to 0$ and $\tau\to 1$. For $\tau\to 0$, one finds
from \re{ab}
\be\label{ab-small}
a=\frac1{\pi m}\lr{1+\frac{\tau}4}+...
\,,\qquad
b=\frac1{\pi m}\lr{1-\frac{\tau}4}+...\,,\qquad\beta=1-\frac{\tau^2}{16}+...\,,
\ee
where the ellipses denote subleading terms. Since $a-b\sim \tau \to 0$, the two
cuts shrink into points as $\beta\to 1$. One expands \re{Energy-2-cut} in powers
of $\tau$ and obtains the energy as
\be\label{E-small}
\varepsilon \stackrel{L \gg N}{=} \frac{m^2}{8L} {\pi^2} \tau^2 + \ldots =
\frac{N}{L^2} \frac{2 \pi^2 m^2}{s}+\ldots \,.
\ee
For $\tau\to 1$, one has from \re{ab}
\be\label{ab-large}
a=\frac1{2m}+...\,,\qquad b=\frac{\sqrt{1-\tau}}{2m}+ ... \,,\qquad
\beta= \sqrt{1-\tau}\,\ln\frac{1}{\sqrt{1-\tau}}+ ...
\ee
According to \re{beta}, the limit $\beta\to 0$ corresponds to $\beta=Ls/N$ with
$N \gg L$. Then, making use of \re{Energy-2-cut} one finds for the energy
\be\label{E-large}
\varepsilon \stackrel{L \ll N}{=} \frac{4m^2}{L} \ln^2 \frac{\sqrt{1-\tau}}{4} +
\ldots = \frac{4m^2}{L} \ln^2 \frac{N}{L}+ \ldots \, .
\ee
This relation matches the expression for the energy of long spinning string
folded $m$ times~\cite{FroTse03} and agrees with the thermodynamic Bethe ansatz
analysis \cite{BeiFroStaTse03}.

\subsection{Breakdown of semiclassical expansion}

For $m=1$ the expressions \re{Energy-2-cut}, \re{E-small} and \re{E-large}
describe the dependence of the energy on the total spin $N+Ls$ along the minimal
energy trajectory at leading order of the semiclassical expansion for $L\gg 1$
and arbitrary values of $N$ including the regions $N\ll L$ and $N\gg L$. {}From
\re{E-naive-expansion} one expects that the contribution of the subleading terms
to the energy is suppressed by powers of $\eta=1/(L+Ns)$. Let us compare these
predictions with the exact expressions for the minimal energy obtained from
numerical solution to the Baxter equation, Eqs.~\re{Baxter-eq} and
\re{Energy-Baxter}. For $L=10$, the comparison is shown in
Fig.~\ref{Fig-roots-energy10} on the right panel. We observe that the curve
determined by \re{Energy-2-cut} agrees with the exact numbers for $N < L$ and
deviates significantly from them for $N > L$. Most importantly, for $N \gg L$ the
exact energy scales as $\sim 4\ln N$ while the semiclassical formula \re{E-large}
yields a different asymptotic behavior.

To understand the reason for this discrepancy, let us revisit the calculation of
the energy in the previous section and pay special attention to the position of
the cuts. By construction, the cuts run along two symmetric intervals
$[-a,-b]\cup[b,a]$. According to \re{ab-large}, for $\tau\to 1$ the parameter $a$
approaches the value $a=1/(2m)$ while the parameter $b$ vanishes indicating that
the two cuts collide at the origin. Then, one examines the expression for the
momentum \re{dp-res} and finds that for $b=0$ the differential $dp$ reduces to
\re{dp-reduced}. We argued in Sect.~\ref{CollisionCuts} that Eq.~\re{dp-reduced}
leads to a new, logarithmic scaling of the energy \re{Emin-naive}. This suggests
that the problem with recovering this regime within the semiclassical approach
does not lie in the construction of the momentum $p(x)$ but rather in the expression
for the energy \re{E-naive-expansion}.

The exact formula for the energy \re{Energy-Baxter} involves $S'(x_\pm)$
evaluated at $x_\pm=\pm i\beta/L$. The relation \re{E-naive-expansion} was
derived under the assumption that $S'(x)$ is analytical at the vicinity of these
points. For the two-cut solution constructed in the previous section this
assumption is justified provided that the inner boundary of the cut $b$ stays
finite in the limit $L\to\infty$. For $x_\pm \to 0$ with $b={\rm fixed}$, the
semiclassical analysis is applicable and one arrives at the asymptotic behavior
\re{E-small} and \re{E-large} depending on the ratio $N/L$. The situation is more
complex when $b$ vanishes in the scaling limit. For $b\to 0$ with $x_\pm ={\rm
fixed}$, the two cuts collide and the very assumption that $S'(x)$ is analytical
at the origin does not hold true anymore. In this case, one expects that the
energy will develop a new asymptotic behavior and the crossover will occur for $b
\sim |x_\pm|$ (see Fig.\ \ref{CollisionCutsFig} (a) and (b)).

Defining a new parameter $\xi =|x_\pm|/b= {\beta}/{(bL)}$ one anticipates that
the semiclassical approach is applicable for $\xi < 1$. Indeed, for $N\ll L$ one
finds from \re{ab-small} that $\xi \sim 1/L$ and the condition  $\xi < 1$ is
satisfied. For $N\gg L$ one gets from \re{ab-small}
\be
\xi \sim \frac1{L} \ln \frac{N}{L}\,.
\ee
One recognizes that similar parameter $\xi_{\rm str}$ naturally arises on the
string side and controls there the transition between two different regimes in
\re{3rd}. For $\xi \gg 1$ one has to take into account that the two cuts collide at the
origin and, as a consequence, the analytical properties of $S'(x)$ are different.
The simplest way to see this is to examine the Taylor expansion of the leading
term $S_0''(x)$, Eq.~\re{SO-zero}, around the origin and retain the terms most
singular for $b\to 0$
\be
S_0''(x) \sim \frac{\beta}{x^2}\left[1 -
\frac1{\sqrt{1-x^2/b^2}}\right]=-\frac{\beta}{2b^2}\left[1+\frac34\,{\frac
{{x}^{2}}{{b}^{2}}}+\frac58\,{\frac {{x}^{4}}{{b}^{4}}}+\ldots
\right]\,.
\ee
As follows from \re{tau-expansion} and \re{E-naive-expansion}, higher order terms
of this expansion contribute to subleading corrections to the energy
$\varepsilon=i\int_{x_-}^{x_+} dx \, S''(x)$. It is easy to see that the series
for $i\int_{x_-}^{x_+} dx \, S_0''(x)$ runs in powers of $\xi=|x_\pm/b|$ and it
is only convergent for $\xi < 1$.

Based on our analysis we expect that the semiclassical expansion of the energy
\re{E-naive-expansion} will be divergent for $\xi \gg 1$. Let us calculate the
first subleading correction to the energy \re{E-naive-expansion} defined by
$S_1''(0)$. The function $S_1'(x)$ satisfies the relation \re{S1-Baxter} which
involves yet unknown function $\tau_1(x)$. To determine $S_1'(x)$ one requires
that it should have the same analytical properties as the leading term of the
semiclassical expansion $S_0'(x)$, i.e., to be an analytical function in the
complex plane with the cuts $[-a,-b]\cup [b,a]$. Discontinuity of $S_1'(x)$
across the cut defines the correction to the distribution density of the Bethe
roots~\cite{Kor95}. Since $\tau_0(x)$ and $\tau_1(x)$ are entire functions while
$p'(x)$ and $\sin p(x)$ change sign across the cut, $\sin p(x+i0)=-\sin p(x-i0)$,
the function $S_1'(x)$ satisfies the relation
\be
S_1'(x+i0) + S_1'(x-i0) = -p'(x+i0) \cot p(x+i0)\,,
\ee
with $x \in [-a,-b]\cup [b,a]$. In addition, its asymptotic behavior at the
origin and infinity is fixed by Eqs.~\re{as-infinity} and \re{as-zero},
\be
S_1'(0)=0\,,\qqqquad S_1'(x)\stackrel{x\to\infty}{\sim} \frac1{x^3}\,.
\ee
The solution to the resulting Riemann-Hilbert problem looks like (c.f.
\cite{GroKaz05})
\be
\label{S1intermediate} S_1'(x) = \frac{x}{\sqrt{(x^2-a^2)(x^2-b^2)}}\int_b^a
\frac{dz}{\pi}\frac{\sqrt{(a^2-z^2)(z^2-b^2)}}{x^2-z^2} p'(z+i0) \cot p(z+i0)\,,
\ee
where `$+i0$' indicates that the momentum is evaluated on the upper sheet of the
Riemann surface. Replacing $p'(z+i0)$ by its expression \re{dp-res}, we obtain
the following representation for the first subleading correction to
\re{E-naive-expansion}
\be
\varepsilon_1 = -\frac{2\beta}{L}S_1''(0) = \frac{2\beta}{\pi ab L}\int_b^a
\frac{dz}{z^2} \left[1-\beta\frac{ab}{z^2}\right] \coth(ip(z+i0)) \,.
\ee
For our purposes we would like to determine the asymptotic behavior of
$\varepsilon_1$ for $N \gg L$, or equivalently $a\to 1/(2m)$, $b\to 0$ and
$\beta\to 0$, Eq.~\re{ab-large}. Changing the integration variable to $z\to b z$,
one finds in the limit $b\to 0$
\be\label{eps1}
\varepsilon_1 =   \frac{2\beta}{\pi ab^2L}\int_1^\infty
\frac{dz}{z^2}\left[1-\frac{\beta a}{b}\frac{1}{z^2}\right] \coth(i  p(b z+i0))=
-\frac{2}{3\pi}\frac{\beta^2}{L b^3} + \ldots\, .
\ee
Here in the second relation we took into account that for $\beta\to 0$ the cuts
collide at the origin and the momentum scales as $p(z+i0) \sim 2im \ln z$ for $z
\to 0$, Eq.~\re{dp-reduced}. Comparing \re{eps1} with a similar relation for the
leading term $\varepsilon_0 \sim \beta^2/(L b^2)$, Eq.~\re{Energy-2-cut}, one
finds that
\be
\frac{\eta \varepsilon_1}{\varepsilon_0} = \mathcal{O}(\eta /b)=\mathcal{O}(\xi)
\, ,
\ee
where $\eta/b=1/(b N)=(2m/s) \ln(N/L)/L$ in the limit $N\gg L$, Eqs.~\re{beta}
and \re{ab-large}.

We conclude that for $N\gg L$ the semiclassical expansion of the energy,
$\varepsilon=\varepsilon_0+\eta\varepsilon_1+\mathcal{O}(\eta^2)$,
Eq.~\re{E-naive-expansion} runs in powers of $\xi$ and it is only convergent for
$\xi <1$. Together with \re{gamma=energy} this leads to the expression for the
one-loop anomalous dimension, Eqs.~\re{gen-exp} and \re{coeff-fun}, with
$c_{1,1}=-2m/(3\pi s)$.

\subsection{Beyond semiclassical expansion}
\label{BeyondSection}

We demonstrated in the previous section that the semiclassical expansion breaks
down for $\xi\gg 1$. The reason for this is that the cuts collide in this limit
and the semiclassical expression for $S(x)$ ceases to be analytical at $x=0$.
Substituting $\beta=0$ into \re{SO-zero}, one finds (for $m=1$)
\be
\label{P-log}
S_0''(x) = p'(x)= -\frac1{x\sqrt{x^2-\frac14}}
\ee
and, therefore, $S_0'(x) \sim \mp 2i \ln x$ for $x\to 0$ on the upper and lower
sheets, respectively. It is easy to see that the corresponding transfer matrix
$\tau_0(x)=2\cos p(x)$ is given by \re{tau-min}. We remind that $\tau_0(x)$ plays
the role of the potential in the Baxter equation \re{Baxter-eq}. The fact that
$\tau_0(x)$ is singular at $x=0$ implies that the semiclassical expansion
\re{WKB-ansatz} breaks down at the origin. In this section, we present an
approach which allows one to construct asymptotic solution to the Baxter equation
\re{Baxter-eq} and evaluate the energy for $N > L$. It is complementary to the
semiclassical approach and takes a full advantage of the above mentioned
singularity of the transfer matrix $\tau_0(x)$ at $x=0$. A detailed account on
this approach can be found in Ref.~\cite{DerKorMan99}.

To begin with, we rewrite the Baxter equation \re{Baxter-eq} as
\be\label{Baxter-phi}
(u+is)^L \phi(u) + \frac{(u-is)^L}{\phi(u-i)} = t_L(u)
\ee
where the notation was introduced for
\be
\phi(u)=\frac{Q(u+i)}{Q(u)}\,.
\ee
To evaluate the energy \re{Energy-Baxter} one has to construct solutions to
\re{Baxter-phi} for $u\sim \pm is$. Notice that the dressing factors $(u\pm
is)^L$ in \re{Baxter-phi} vanish for $u\to \mp is$ indicating that one of the
terms in the left-hand side of \re{Baxter-phi} becomes anomalously small and can
be neglected. It turns out that in the thermodynamic limit, $L \gg 1$, the same
approximation can be performed not only in the vicinity of $u=\pm is$ but in the
whole region $u=\mathcal{O}\left((N+Ls)^0\right)$. To see this we note that
$t_L(u)$ is given by \re{t_L} with the charges that scale as $q_k \sim (N+Ls)^k$.
This suggests that $|t_L(u)| \gg 1$ for $u=\mathcal{O}\left((N+Ls)^0\right)$.
Indeed, in the semiclassical approach, for $u=x/\eta$, one finds from
\re{tau-expansion}, \re{p-def} and \re{P-log} that the transfer matrix takes the
form
\be\label{t-p}
t_L(x/\eta) = 2 (x/\eta)^L \cos p(x) + \mathcal{O}(\eta)\,.
\ee
For $u=\mathcal{O}(L^0)$, or equivalently $x\sim \eta$, the momentum can be
replaced by its asymptotic behavior at the origin: $p(x) \sim\beta/x$ for
$\xi\ll 1$ and $p(x)\sim 2i\ln x$ for $\xi \gg 1$ leading to $|t_L(x/\eta)|\gg
1$.

Let us return to Eq.\ \re{Baxter-phi} and take into account that $t_L(u)$ takes
large values for $u=\mathcal{O}\left((N+Ls)^0\right)$ both for $\xi< 1$ and
$\xi\gg 1$. Requiring the left-hand side of \re{Baxter-phi} to be as large as
$t_L(u)$ one finds that either $\phi(u) \gg 1$, or $\phi(u-i) \ll 1$. In both
cases, one of the terms in the left-hand side of \re{Baxter-phi} can be safely
neglected and one arrives at two different equations
\ba\label{Baxter-as}
(u+is)^L Q_+(u+i) \!\!\!&=&\!\!\! t_L(u) Q_+(u)\,,
\\ \nonumber
(u-is)^L Q_-(u-i) \!\!\!&=&\!\!\! t_L(u) Q_-(u)\,.
\ea
Having solved this system, one can construct an asymptotic solution to the Baxter
equation \re{Baxter-eq} as a linear combination of $Q_+(u)$ and $Q_-(u)$
\be\label{Q-as}
Q^{\rm (as)}(u) = c_+ Q_+(u) + c_- Q_-(u)\,.
\ee
We would like to stress that $Q^{\rm (as)}(u)$ does not satisfy the Baxter
equation \re{Q-polynom}, but asymptotically approaches its solution $Q(u)$ in the
region $u=\mathcal{O}\left((N+Ls)^0\right)$. Equations~\re{Baxter-as} were
obtained under the assumption that $(u+is)^L Q_+(u+i)\gg (u-is)^L Q_+(u-i)$ and
$(u+is)^L Q_-(u+i)\ll (u-is)^L Q_-(u-i)$, respectively. Together with
\re{Baxter-as} it can be expressed as the following relation for the transfer
matrix
\be
t_L\left(u+\ft{i}2\right) t_L\left(u-\ft{i}2\right) \gg \left[u^2 +
\left(\ft12-s\right)^2\right]^L\,.
\ee
For $u=\mathcal{O}\left((N+Ls)^0\right)$ it is equivalent to $|t_L(u)|\gg 1$.

To solve \re{Baxter-as} one introduces into consideration the roots of the
transfer matrix \re{t_L}
\be\label{t-roots}
t_L(u) = 2 \prod_{n=1}^L (u-\delta_n)\,.
\ee
It is known that for polynomial solutions to the Baxter equation,
Eq.~\re{Q-polynom}, the roots take real values only, $\Im
\delta_k=0$~\cite{Kor95}. Matching \re{t-roots} into \re{t_L} one finds that they
satisfy the sum rules
\be\label{sum_rules}
\sum_{n=1}^L \delta_n = 0 \,,\qquad \sum_{n=1}^L \delta_n^2 = -\frac12q_2
\,,\quad \ldots\, , \quad \prod_{n=1}^L \delta_n = \frac{(-1)^L}{2}q_L \,.
\ee
For even solutions to the Baxter equation, $Q(u)=Q(-u)$, the roots appear in
pairs $\delta_n = -\delta_{L - n + 1}$. Making use of \re{t-roots}, it is
straightforward to verify that the solutions to \re{Baxter-as} are given by
\ba\label{Q+-}
Q_+(u) = 2^{-iu} \prod_{n=1}^L\frac{\Gamma(-iu + i\delta_n)}{\Gamma(-iu +s )} \,
, \qquad Q_-(u) = 2^{iu} \prod_{n=1}^L\frac{\Gamma(iu - i\delta_n)}{\Gamma(iu +s
)} \, .
\ea
To fix the constants $c_\pm$ in \re{Q-as} one examines the relation for the
quasimomentum in Eq.\ \re{Energy-Baxter} and substitutes $Q(u)$ by its asymptotic
expression \re{Q-as}. Taking into account that (for $\varrho \to 0$)
\be\label{Q-vanish}
Q_+(-is + \varrho) \sim \varrho^L \,,\qquad Q_-(is - \varrho) \sim \varrho^L
\ee
one finds from \re{Energy-Baxter} and \re{Q-as}
\be
\e^{i\theta} = \frac{Q^{\rm (as)}(is)}{Q^{\rm (as)}(-is)}
=\frac{c_+}{c_-}\frac{Q_+(is)}{Q_-(-is)}\,.
\ee
Therefore, for cyclically symmetric states $\e^{i\theta}=1$ the asymptotic
solution to the Baxter equation is given up to an overall normalization factor by
\be\label{Q-as-sol}
Q^{\rm (as)}(u) =  Q_+(u)Q_-(-is) +  Q_-(u)Q_+(is)\,.
\ee
We remind that this relation is only valid in the region
$u=\mathcal{O}\left((N+Ls)^0\right)$. In a similar manner, one uses
\re{Q-vanish} to evaluate the energy \re{Energy-Baxter} as
\be
\varepsilon^{\rm (as)} = i\lr{\ln Q^{\rm (as)}(is)}'-i\lr{\ln Q^{\rm (as)}(-is)}'
= i\lr{\ln Q_+(is)}'-i\lr{\ln Q_-(-is)}' \, .
\ee
Replacing $Q_\pm(u)$ by their expressions \re{Q+-} one finds the following
remarkable expression for the energy in terms of the roots of the transfer matrix
\re{t-roots}~\cite{Kor95,BraDerMan98,Bel99,DerKorMan99}
\be\label{energy-as}
\varepsilon^{\rm (as)} = 2 \ln 2 + \sum_{n=1}^L \left[\psi(s+
i\delta_n)+\psi(s-i\delta_n) -2 \psi(2s) \right] \,,
\ee
where $\psi(x)=d\ln\Gamma(x)/dx$. The interpretation of \re{energy-as} in terms
of classical $SL(2)$ spin magnet and properties of the Wilson operators \re{O-def}
in gauge theory can be found in Ref.~\cite{BelGorKor03}.

Since the roots $\delta_n$ are functions of the conserved charges,
Eq.~\re{sum_rules}, the relation \re{energy-as} establishes the dependence of the
energy on $q_2,\ldots,q_L$. To check \re{energy-as}, we compared the asymptotic
``dispersion curve'' $\varepsilon^{\rm (as)}(q_2,\ldots,q_L)$ with the exact one
$\varepsilon^{\rm (ex)}(q_2,\ldots,q_L)$ coming from the Bethe Ansatz solution
\re{Energy-ABA}. We found that for $L=10$ and $s=1/2$ the accuracy of
\re{energy-as}, $\varepsilon^{\rm (as)}/\varepsilon^{\rm (ex)}-1 $, increases
from $-4.6 \times 10^{-5}$ for $N=2$, to $-2.6 \times 10^{-6}$ for $N=10$ and to
$1.9 \times 10^{-10}$ for $N=100$. We conclude that Eq.~\re{energy-as} describes
the exact eigenspectrum with a high accuracy throughout the whole interval of $N$
including the region $N\sim L$ in which the semiclassical approach is applicable.
Indeed, it is straightforward to show that the relation \re{energy-as} coincides
with the semiclassical expression \re{E-naive-expansion} for $N\sim L$ (see
Appendix).

The charges $q_3,\ldots,q_L$ have to satisfy quantization conditions. In the
method of Baxter $Q-$operator they follow from the requirement for $Q(u)$ to be
polynomial solutions to the Baxter equation, Eqs.~\re{Baxter-eq} and
\re{Q-polynom}. The roots of $Q(u)$ scale in the thermodynamic limit as
$u=\lambda_j \sim L$ and, therefore, they lie outside the applicability range of
\re{Q-as-sol}. This does not allow us to impose the polynomiality condition for
$Q^{\rm (as)}(u)$. We shall require instead that $Q^{\rm (as)}(u)$, given by
\re{Q-as-sol}, should be regular on the real $u-$axis inside the region
$u=\mathcal{O}\left((N+Ls)^0\right)$. This property is not warranted since both
$Q_+(u)$ and $Q_-(u)$ develop poles on the real $u-$axis originating from the
product of $\Gamma-$functions in the numerator in the right-hand side of
Eq.~\re{Q+-}. The poles are located at zeros of the transfer matrix \re{t-roots},
$u=\delta_j$ with $j=1,\ldots,L$. Requiring for $Q^{\rm (as)}(u)$ to have zero
residues at $u=\delta_n$ with $\delta_n=\mathcal{O} \left((N+Ls)^0\right)$, one
gets from \re{Q-as-sol} and \re{Q+-}
\be\label{quant-cond}
2^{-2i\delta_n} \prod_{j=1, j\neq n}^L
\frac{\Gamma(-i\delta_n+i\delta_j)}{\Gamma(i\delta_n-i\delta_j)}=
\left[\frac{\Gamma(s-i\delta_n)}{\Gamma(s+i\delta_n)}\right]^L \prod_{j=1}^L
\frac{\Gamma(s+i\delta_j)}{\Gamma(s-i\delta_j)}\,.
\ee
The total number of roots of the transfer matrix \re{t-roots} equals the length
of the spin chain $L$. As we will see in a moment, in the thermodynamic limit $L
\gg 1$ and $N \gg L$ all roots can be separated into two different groups
depending on their scaling: ``small'' roots $\delta_n=\mathcal{O}(N^0)$ and
``large'' roots $\delta_n=\mathcal{O}(N)$. It is important to realize that the
quantization conditions \re{quant-cond} should only hold for the small roots
while the product over $j$ entering both sides of \re{quant-cond} involves {\sl
all} roots.

The total number of small roots depends on the value of the parameter
$\beta=Ls/(N+Ls)$. For $0 < \beta \le 1$, that is, within the applicability range
of the semiclassical expansion, the integrals of motion scale as in
Eq.~\re{q-scaling}. Together with \re{sum_rules} this implies that the roots
behave as $\delta_n = \mathcal{O}(N)$ and, therefore, there are no small roots.
For $\beta\to 0$ higher charges $q_L, q_{L-1}, \ldots $ take anomalously small
values \re{q-scaling} indicating that the small roots are there. Let us
demonstrate that their total number equals $L-2m$ with positive integer $m$
entering the right-hand side of \re{cond}. One makes use of \re{t-p} to deduce
that the large roots of the transfer matrix satisfy $\cos p(\delta_j\eta)=0 +
\mathcal{O}(\eta)$. Recall that the momentum verifies $(\cos p(x))^2 \ge 1$ on
the cuts (for $b^2 \le x^2 \le a^2$) and takes the values $\cos p(x)=\pm 1$ at
the double points $x=x_{2j}$, Eq.~\re{equid}. Since $\cos p(x_{2j})=-\cos
p(x_{2j-2})$, the roots of the transfer matrix \re{t-p} should lie on the real
axis in between the branching points, $x_{2k} < \delta_j\eta < x_{2j-2}$, away
from the cuts $[-a,-b]\cup [b,a]$ (see Fig.~\ref{Fig-semi}). Going over to the
limit $\beta =Ls/(N+Ls)\to 0$, or equivalently $N \gg L$, one takes into account
that the edges of the cuts scale as $a = 1/{(2m)}$ and $b=\beta/(-2m\ln\beta)\to
0$, Eq.~\re{ab-large}. As a result, all roots of the transfer matrix
$\delta_j\eta\sim \delta_j/N$ ``trapped'' inside the interval $[-b,b]$ are to be
small while the roots belonging to the intervals $(-\infty,-a]\cup [a,\infty)$
are to be large
\be
\label{roots-as} |\delta^{\rm (large)}| > \frac{N}{2m} \,,\qqqquad |\delta^{\rm
(small)}| < \frac{1}{4m\xi}\,,
\ee
with $\xi=\ln (N/L)/L$. Thus, the total number of small roots, $L-2m$, equals the
number of double points on the interval $[-b,b]$ and, as a consequence, the total
number of the large roots is $2m$.

Let us consider the minimal energy state with $m=1$. For $N \gg L$, the transfer
matrix has two large roots $\delta_1=-\delta_{L}$ and $L-2$ small roots
$\delta_k=-\delta_{L-k+1}$ with $k=2,\ldots,L-1$. It follows from \re{sum_rules}
that the large root is given by
\be\label{delta_1}
\delta_1 = {(-q_2/2)^{1/2}} \left[ 1+ \mathcal{O}(1/L)\right]
\sim
\frac{N+Ls}{\sqrt{2}}
\ee
with $q_2$ defined in \re{q2}. The small roots satisfy the quantization
conditions \re{quant-cond}. Separating in \re{quant-cond} the contribution of the
large roots with $j=1$ and $j=L$, one can rewrite \re{quant-cond} in terms of the
small roots only (for $n=2,\ldots,L-1$)
\be\label{system}
(-q_2)^{-2i\delta_n} = (-1)^{L-1}
\left[\frac{\Gamma(s-i\delta_n)}{\Gamma(s+i\delta_n)}
\right]^{L}\prod_{j=2}^{L-1}
\frac{\Gamma(1+i\delta_n-i\delta_j)}{\Gamma(1-i\delta_n+i\delta_j)} \,.
\ee
Here we took into account that the product over $j$ in the right-hand side of
\re{quant-cond} equals $1$ due to pairing of the roots
$\delta_n=-\delta_{L-n+1}$. Taking the logarithm of both sides in \re{system},
one rewrites the quantization conditions as
\be\label{system-arg}
\delta_n \ln(-q_2) + L \,{\rm arg}\, \Gamma(s-i\delta_n) + \sum_{j=2}^{L-1} {\rm
arg}\, \Gamma(1+i\delta_n-i\delta_j) = \frac{\pi}2 k_n
\ee
where integers $k_2> k_3> \ldots$ count the branches of the logarithms and
satisfy $k_n=-k_{L-n+1}$. In addition, they have a parity opposite to that of
$L$, that is $k_n={\rm even/odd}$ for $L={\rm odd/even}$. Notice that in
distinction with the Bethe ansatz equations \re{Bethe-roots}, the number of
relations in \re{system-arg} does not depend on the total spin $N$ but only on
the length of the spin chain $L$.

To evaluate the energy \re{energy-as} one has to solve the system \re{system-arg}
and determine the set of small roots. The resulting expression for
$\varepsilon^{\rm (as)}$, Eq.~\re{energy-as}, depends on integers $k_n$ and as we
show below (see Eq.~\re{Ek}) it takes minimal value for the occupation numbers
$k_n=L+1-2n$. A systematic analysis of the quantization conditions
\re{system-arg} will be given elsewhere. The system of equations \re{system-arg}
can be easily solved in two limits $\xi \gg 1$ and $\xi < 1$.

For $\xi \gg 1$ (or $\ln(N/L) \gg L $) one deduces from \re{roots-as} that
solutions to \re{system-arg} have to satisfy $|\delta_n| \ll 1$. In this case,
one expands the $\Gamma-$functions in \re{system-arg} in powers of $\delta$'s and
one finds after some algebra
\be\label{delta-fin}
\delta_n = \frac{\pi k_n/2}{\ln(-q_2) + (L-2)\psi(1)-L\psi(s)}+\ldots
\ee
with $n=2,\ldots,L-1$. One verifies a posteriori that this relation is in
agreement with \re{roots-as}.

For $\xi < 1$ (or $1 \ll \ln(N/L) <  L $), one obtains from \re{roots-as} that
$|\delta_n| < 1/(2\xi)$ and, therefore, not all roots verify the relation
$|\delta_n| \ll 1$. Still, the relation \re{delta-fin} is valid for the roots
with small absolute value $|\delta_n| \ll 1$ (with $n\sim L/2$). For roots
$|\delta_n| \sim 1/(2\xi)$ (with $n \sim L^0$) one expands the $\Gamma-$functions
in \re{system-arg} in inverse powers of $\delta$'s and obtains
\be\label{dev}
\delta_n\ln(-q_2) -  \delta_n \ln \delta_n^2 +\ldots  =\frac{\pi}{2} k_n\,,
\ee
with $k_n = \mathcal{O}(L)$. From this relation one obtains the relation
$\delta_n \sim \pi k_n/(4 \ln(N/k_n))$ which is in agreement with \re{roots-as}.

\begin{figure}[t]
\begin{center}
\mbox{
\begin{picture}(0,230)(230,0)
\put(-10,0){\insertfig{8}{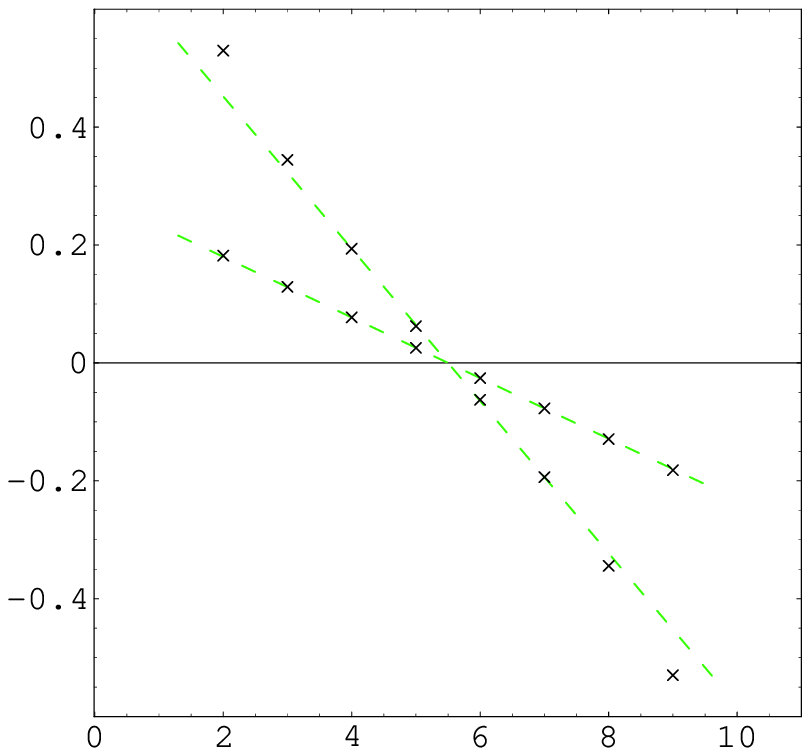}} \put(-10,116){$\delta_n$}
\put(110,-5){$n$} \put(180,50){$\scriptstyle{N = 10^2}$}
\put(180,100){$\scriptstyle{N = 10^{10}}$}
\put(250,5){\insertfig{7.65}{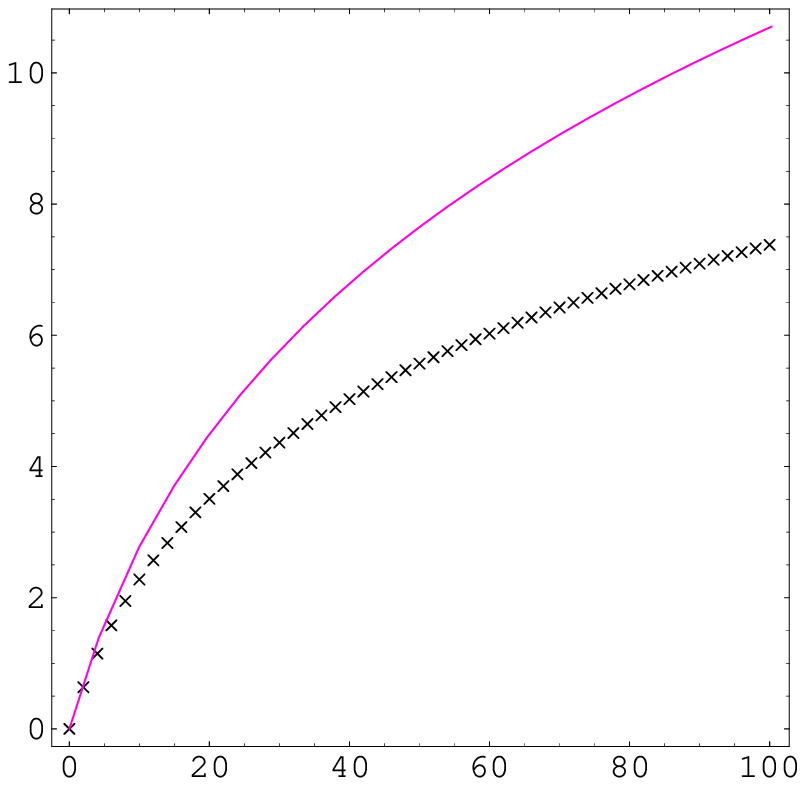}} \put(245,116){$\varepsilon$}
\put(360,-5){$N$}
\end{picture}
}
\end{center}
\caption{ \label{Fig-roots-energy10} Left panel: ``small'' roots of the transfer
matrix $\delta_n$ (with $n=2,\ldots,L-1$) for $s=\ft12$, $L=10$ and two values of
the spin $N = 10^2$ and $N=10^{10}$. Crosses stand for the solutions to the
quantization condition \re{system-arg} and the lines correspond to \re{delta-fin}
for $k_n=L+1-2n$. The roots tend to approach the line as $\xi=\ln(N/L)/L$
increases from $\xi = 0.23$ to $\xi=2.07$. Right panel: the minimal energy for
$s=\ft12$, $L=10$ and the total spin $0 \le N \le 100$. Crosses denote the exact
values, Eq.~\re{Energy-ABA}, while the solid line stands for the semiclassical
expression \re{Energy-2-cut}. We do not display the data for the asymptotic
energy \re{energy-as} with roots deduced from the quantization conditions
\re{system-arg}, since they are not distinguishable from the exact spectrum.}
\end{figure}

To test the quantization conditions \re{system-arg} we compared solutions to
\re{system-arg} for $k_n=L+1-2n$ with the exact, Bethe ansatz expressions for the
roots of the transfer matrix corresponding to the minimal energy state with the
quantum numbers $s=\ft 12$, $L=10$ and $N=100$ and observed a good agreement (see
Table \ref{tab:WKB}).
%
\begin{table}[t]
\begin{center}
\begin{tabular}{|c||c|c|c|c|c||c|}
\hline  & $\delta_1$ & $\delta_2$ &  $\delta_3$ &
     $\delta_4$ &  $\delta_5$ &
     $E$
\\
\hline exact & $73.897289$  & $0.5297596$ & $0.3443964$ & $0.1937491$ &
$0.06253897$ & $7.3790455$
\\
\hline asym & $73.900271$ & $0.5297573$ & $0.3443955$ & $0.1937487$ &
$0.06253886$ & $7.3791719$
\\
\hline
\end{tabular}
\end{center}
\caption{Comparison of the exact roots, $\delta_n=-\delta_{L-n+1}$, and the
energy, $E$, for $s=1/2$, $L=10$ and $N=100$ with the asymptotic expressions
obtained from Eqs.~\re{system} and \re{energy-as}, respectively.} \label{tab:WKB}
\end{table}
%
In agreement with \re{delta-fin} and \re{dev}, the small roots $\delta_n$ vary
linearly with $n$ close to the origin and deviate from the linear behavior close
to the end points. Moreover, for $\xi \gg 1$ all small roots scale linearly with
$n$ in agreement with \re{delta-fin} (see Fig.~\ref{Fig-roots-energy10}).

In a similar manner, we solved quantization conditions \re{system-arg} for
$s=\ft12$, $L=10$ and $0 \le N \le 100$ and compared the resulting expression for
the energy \re{energy-as} with the exact expression \re{Energy-ABA} as shown in
Table~\ref{Fig:energy}.
%
\begin{table}[t]
\begin{center}
\begin{tabular}{|c||c|c|c|c|c|c|c|c|c|c|}
\hline $N$ & $0$ & $10$ &  $20$ &
     $30$ &  $40$ &
     $50$ & $60$ & $70$ & $80$ & $90$
\\
\hline exact & $0$  & $2.2766$ & $3.5069$ & $4.3644$ & $5.0266$ & $5.5678$ &
$6.0262$ & $6.4243$ & $6.7765$ & $ 7.0924$
\\
\hline asym & $0.0907$ & $2.2851$ & $3.5097$ & $4.3657$ & $5.0274$ & $5.5683$ &
$6.0265$ & $6.4246$ & $6.7767$ & $7.0926$
\\
\hline
\end{tabular}
\end{center}
\caption{Comparison of the exact energy, $\varepsilon$, with the asymptotic
expression obtained from Eqs.~\re{system} and \re{energy-as} for $s=\ft12$,
$L=10$ and different total spin $N$.} \label{Fig:energy}
\end{table}%
We found that for $N \ge L$ the asymptotic expression \re{energy-as} approximates
the exact result with an accuracy better than $0.002\%$ while the semiclassical
approach significantly overestimates the value of the energy (see
Fig.~\ref{Fig-roots-energy10}). Remember that the minimal energy states carry
even Lorentz spin only, Eq.~\re{q-odd}.

Let us examine the expression for the energy \re{energy-as} in the limit $\xi \gg
1$, or $\ln(N/L) \gg L $. We remind that the semiclassical expansion breaks down
in this region. Equation~\re{energy-as} involves the sum over the roots of the
transfer matrix \re{t-roots}. As before, we separate them into two groups.
According to \re{roots-as}, the large and small roots scale as $\delta_j \sim N$
and $\delta_j\sim 1/\xi$, respectively. This allows one to replace $\psi(s\pm
i\delta_j)$ in \re{energy-as} by its asymptotic behavior at infinity and at the
origin, respectively. In this way we obtain 
\be
\varepsilon = 2\ln 2 - 2L \psi(2s)+ 2 \sum_{\rm large}  \ln |\delta_j|  +
\sum_{\rm small} \left[2\psi(s) - \psi''(s)\,\delta_j^2\right] + \ldots\,,
\ee
where the ellipsis denotes subleading terms. The number of large roots equals $2
m$, $\delta_j=-\delta_{L-j+1}$ (with $j=1,\ldots,m$) and, therefore, the leading
asymptotic behavior of the energy for $\xi\gg 1$ is~\cite{BelGorKor03}
\be
\varepsilon = 4 \ln (\delta_1 \delta_2 \ldots \delta_m) + \mathcal{O}(N^0) = 2
\ln |q_{2m}| + \mathcal{O}(N^0)\,,
\ee
where we took into account the relations \re{sum_rules} and $q_{2m}= \widehat
q_{2m}^{\scriptscriptstyle (0)} N^{2m}$ was defined in \re{q-expression}. In this
way we obtain the relation
\be\label{SingleLogAsymptotics}
\varepsilon = 4 m \ln N + \mathcal{O}(N^0)\,,
\ee
which is in an agreement with \re{E-m}. The minimal energy corresponds to $m=1$.

For $m=1$ one has just two large roots $\delta_1=-\delta_{L}$ and $L-2$ small
roots $\delta_n=-\delta_{L-n+1}$ (with $n=2,\ldots,L-1$) defined in
Eqs.~\re{delta_1} and \re{delta-fin}, respectively. Therefore, for $\xi \gg 1$
one gets
\be\label{Ek}
\varepsilon = 4\ln N  + 2 L \left[ \psi(s)-\psi(2s)\right] + \frac{ (-\psi''(s))
\,\pi^2 }{16\ln^2 N }\sum_{n=2}^{L-1}{k_n^2} +\ldots
\ee
This expression depends on the set of integers $k_2> k_3> \ldots$ (with
$k_n=-k_{L-n+1}$) defined in \re{system-arg}. Since $\psi''(s)<0$ for $s>0$, the
minimal value of $\varepsilon$ corresponds to $k_n=L+1-2n$
\be\label{e-minimal}
\varepsilon = 4\ln N  + 2 L \left[ \psi(s)-\psi(2s)\right] + \frac{L^3
(-\psi''(s)) \,\pi^2}{48\ln^2 N } +\ldots
\ee
This relation defines the minimal energy for $\xi \gg 1$. The energy of the
excited states is described by \re{Ek} with another set of $k-$integers but of
the same parity. The lowest lying excited state has the same $k$'s as the minimal
energy state except $k_2=-k_{L-1}=L-1$. It is separated from the latter by the
distance
\be
\Delta \varepsilon = \frac{ \pi^2 L (-\psi''(s))}{2\ln^2 N}\sim \frac{L}{\ln^2 N}
\,.
\ee
This relation defines the level spacing in the spectrum of anomalous dimension of
the Wilson operators \re{O-def} close to the minimal anomalous dimension
trajectory.

So far our discussion was limited to one loop. Remarkably enough the logarithmic
behavior of the anomalous dimension \re{gamma=energy}  persists to all orders of
perturbation theory. Higher order corrections to \re{gamma=energy} merely modify
the coefficient in front of $\ln N$ replacing $\lambda$ by an infinite series in
the coupling constant,
\be
\gamma(\lambda) = \frac{\lambda}{8\pi^2}\, \left[4\ln N + \mathcal{O}(N^0)\right]
+ \mathcal{O}(\lambda^2)= 2\Gamma_{\rm cusp} (\lambda) \ln N +
\mathcal{O}(N^0)\,,
\ee
with $\Gamma_{\rm cusp} (\lambda)$ being the cusp anomalous dimension. This
relation holds true in a generic Yang-Mills theory for arbitrary values of the
coupling constant $\lambda$ including the strong coupling regime. In the latter
case, one can apply the gauge/string correspondence to obtain a prediction for
the cusp anomalous dimension at strong coupling in the $\mathcal{N}=4$ SYM
theory. In the next section, we shall trace the origin of logarithmic scaling of
anomalous dimension in the dual picture of folded string rotating on the AdS
space.

\section{Classical strings in AdS${}_3 \times$S$^1$}
\label{StringSigmaModel}

Let us turn to the analysis of anomalous dimensions of composite operators
\re{O-def} on the string side. According to the gauge/string correspondence
the Wilson operators are mapped into certain string states whose energy is
identified with the scaling dimensions of the former. It is known that in
the $\mathcal{N}=4$ SYM theory, Wilson operators with the minimal anomalous
dimension discussed in the preceding sections are dual to a single-folded
string rotating in the AdS$_3 \times$S$^1$ sector of the target space of the
type IIB string theory \cite{GubKlePol03,FroTse03}.

\subsection{Folded rotating string}

For the folded closed string spinning both in AdS${}_3$ and S${}^1$, the
relevant bosonic part of the superstring action reads
\be\label{NG}
S = - \frac{\sqrt{\lambda}}{2\pi} \int d^2 \xi \sqrt{- \det\| G_{MN}(X)\,
\partial_a X^M \partial_b X^N\|} \, ,
\ee
where the 't Hooft coupling $\lambda=g^2 N_c$ is related to the product of the
radius of anti-de Sitter space $R$ and the string tension $1/\alpha^\prime$,
$\sqrt{\lambda} = R^2/\alpha^\prime$, and the metric in the target space looks
like
\be
ds^2 \equiv G_{MN} d X^M d X^N = - \cosh^2 \rho \, dt^2 + \sinh^2 \rho\,
d\theta^2 + d\rho^2 + d\varphi^2 \, .
\ee
Here $t$, $\rho$ are $\theta$ are the global time, the radial coordinate and
the angle, respectively, on the AdS${}_3$ space and $\varphi$ is the angle on
a large circle of S${}^5$. The worldsheet coordinates of the string  $\xi^a =
(\xi^0,\xi^1)$ are chosen in such a way that
\be\label{para}
t = \xi^0 \, , \qquad \theta = \omega \xi^0 + \xi^1 \, , \qquad \varphi = \nu
\xi^0 \, .
\ee
Here $\omega$ is the angular velocity on AdS and $\nu$ is a boost parameter of
the center-of-mass of the folded string on S${}^5$. For the rigid folded string
the radial variable $\rho$ does not depend on $\xi^0$ and is periodic in $\xi^1$,
that is, $\rho (\xi^1 + 2 \pi) = \rho (\xi^1)$. The dependence $\rho=\rho(\xi^1)$
is determined by the classical string equations of motion
\cite{GubKlePol03,FroTse03,Kru05}. Their solution describes a folded string
configuration which is sewed out of four segments running along the radial
direction from the center of AdS space, $\rho=0$, towards its boundary by a
distance $\rho_0$, which depends on the angular velocities entering \re{para}
\be
\coth^2 \rho_0 = \frac{\omega^2 - \nu^2}{1 - \nu^2}\ge 1\,.
\ee
The string has two spikes located at $\rho=\rho_0$ which are responsible for the
logarithmic scaling of the anomalous dimension of Wilson operators at strong
coupling~\cite{BelGorKor03,Kru05}.

Within the gauge/string correspondence, the anomalous dimension of the operators
\re{O-def} is related to the energy $E$ of the classical rotating string as
\be\label{E-gamma}
E = N + L + \gamma(\lambda)
\, ,
\ee
where the Lorentz spin $N$ and twist $L$ are translated into the angular momenta
on the AdS$_3$ and S$^1$ spaces, respectively. From the Nambu-Goto action
\re{NG}, one finds these charges as
\be\label{Efolded}
E = \frac{2 \sqrt{\lambda}}{\pi} \frac{\sqrt{- \chi}}{\sqrt{1 - \nu^2}}
\mathbb{E} (\chi)\,,\qquad N=\frac{2 \sqrt{\lambda}}{\pi} \frac{\omega \sqrt{-
\chi}}{\sqrt{1 - \nu^2}} \left[ \mathbb{E} (\chi) - \mathbb{K} (\chi) \right] \,
,\qquad L=\frac{2 \sqrt{\lambda}}{\pi} \frac{\nu \sqrt{- \chi}}{\sqrt{1 - \nu^2}}
\mathbb{K} (\chi)\,,
\ee
where $\mathbb{K}(\chi)$ and $\mathbb{E}(\chi)$ are the elliptic functions of
the first and second kind, respectively, and the auxiliary (negative valued)
parameter $\chi$ is related to the distance $\rho_0$ from the center of anti-de
Sitter space to the end of the string extending to its boundary,
\be\label{rho0}
\chi = -\sinh^2 \rho_0 \,.
\ee
Excluding $\omega$ and $\nu$ in favor of $\chi$, Eqs.~\re{Efolded} can be cast
into the parametric form \cite{BeiFroStaTse03}
\be
\label{FrolovTseytlin} \left( \frac{E}{\mathbb{E} (\chi)} \right)^2 - \left(
\frac{L}{\mathbb{K} (\chi)} \right)^2 = - \frac{4\lambda}{\pi^2}\chi \, , \qquad
\left( \frac{N}{\mathbb{E} (\chi) - \mathbb{K} (\chi)} \right)^2 - \left(
\frac{L}{\mathbb{K} (\chi)} \right)^2 = \frac{4\lambda}{\pi^2} ( 1 - \chi ) \, .
\ee
Using these relations, one can analyze the anomalous dimension $\gamma=E-N-L$ at
strong coupling in the semiclassical limit of large angular momenta $L$, $N \to
\infty$ with $N/\sqrt{\lambda}$ and $L/\sqrt{\lambda}$ kept fixed. In this way
one finds from \re{FrolovTseytlin} that the two limiting cases $N \ll L$ and $N
\gg L$ correspond to the short and long strings, $\rho_0\ll 1$ and $\rho_0\gg 1$,
respectively.

It is well-known~\cite{GubKlePol03,FroTse03} that in the short string limit,
$\rho_0\ll 1$, or equivalently $(-\chi)\ll 1$, the anomalous dimension exhibits
the BMN scaling
\be
\label{AnomalousDimensionPertExp} \gamma (\lambda) = L\left[ \lambda'
\gamma^{(0)}+ {(\lambda')}^2\gamma^{(1)}+ \ldots \right] \, ,
\ee
with $\gamma^{(0)}, \gamma^{(1)},\ldots$ being functions of the ratio $N/L$ and
$\lambda'=\lambda/(\pi L)^2$ determining the BMN coupling constant. Substituting
$\chi = \chi_0 + \lambda^\prime \chi_1 + \dots$ into \re{FrolovTseytlin} and
matching the coefficients in the expansion of both sides of \re{FrolovTseytlin}
in powers of $\lambda'$, one finds for $N/L \ll
1$~\cite{BerMalNas02,FroTse03,BeiFroStaTse03}
\be
\gamma^{(0)}=\frac{\pi^2}{2}\frac{N}{L}+\ldots \,,\qquad
\gamma^{(1)}=-\frac{\pi^4}{8}\frac{N}{L}+\ldots\,.
\ee
The lowest order term in \re{AnomalousDimensionPertExp} matches the one-loop
result on the gauge theory side, Eq.~\re{E-small}, for $s=\ft12$ and $m=1$
\be\label{match}
\gamma^{(0)} = \frac{L}{8} \varepsilon\,.
\ee
In a similar manner, in the long string limit, $\rho_0 \to \infty$, or
equivalently $(-\chi)\to\infty$, assuming that $\chi$ has a regular expansion in
powers of $\lambda'$, one finds from Eq.\ \re{FrolovTseytlin} that the leading
order parameter $\chi_0$ scales for $N/L\gg 1$ as
\be
\chi_0 = - \frac{N}{2 L} \ln \frac{N}{L} \, .
\ee
This leads to the following expressions for the coefficients of the BMN expansion
of the anomalous dimension \re{AnomalousDimensionPertExp} for $N/L\gg 1$
\ba
\label{AsyPertExp} \gamma^{(0)} \!\!\!&=&\!\!\! \ft12 \ln^2 (- \chi_0)+\ldots
=\ft12 \ln^2 (N/L)+\ldots\, , \\ \nonumber \gamma^{(1)} \!\!\!&=&\!\!\!
-\ft{1}{8} \ln^4 (-  \chi_0) +\ldots = -\ft18 \ln^4 (N/L)+\ldots\, ,
\ea
where ellipses denote terms subleading for $(-\chi_0)\to\infty$. In agreement
with \re{match}, the relation \re{AsyPertExp} matches the one-loop expression for
the anomalous dimension on the gauge theory side, Eq.~\re{E-large}. Moreover, one
can show~\cite{BeiFroStaTse03} that the relation \re{match} (upon the Landen
transformation defined below in Eq.\ \re{LandenGauss}) holds for an arbitrary
value of $N/L$ with $\gamma^{(0)}$ given by \re{FrolovTseytlin} and
\re{AnomalousDimensionPertExp} and $\varepsilon$ defined in \re{Energy-2-cut} and
\re{ab}.

So far we see no trace of expected scaling of the anomalous dimension $\gamma
(\lambda) \sim \ln N$ for $N\gg L$. We recall that on the gauge theory side the
semiclassical expansion for the one-loop energy $\varepsilon$ was divergent for
$N\gg L$ and in order to recover the logarithmic scaling $\varepsilon\sim \ln N$
we had to resum the entire semiclassical series in powers of the parameter
$\xi=\ln(N/L)/L$. It turns out that a similar phenomenon happens for the series
\re{AnomalousDimensionPertExp} on the string side although the parameter of the
semiclassical expansion is different and equals
\be\label{xi-str}
\xi_{\rm str}=\frac{\lambda}{L^2} \ln^2\frac{N}{L}=\lambda\, \xi^2\,.
\ee
It can be easily identified by comparing the contribution to
\re{AnomalousDimensionPertExp} from $\gamma^{(0)}$ and $\gamma^{(1)}$,
Eq.~\re{AsyPertExp}.

To sum up the infinite series in \re{AnomalousDimensionPertExp} for $\xi_{\rm
str}\gg 1$ we return to \re{FrolovTseytlin} and rewrite the first relation as
\be\label{gamma-sqrt}
\gamma(\lambda) = L\left[\sqrt{1- {4\lambda'} {\chi}{\mathbb{K}^2(\chi)}}-1+
\Delta\gamma\right]
\ee
with $\lambda'=\lambda/(\pi L)^2$ and
\be
\Delta\gamma=-
\frac{{4\lambda'}\,\mathbb{K}(\chi)[\mathbb{E}(\chi)-\mathbb{K}(\chi)]}{\sqrt{1-{4\lambda'}
\chi {\mathbb{K}^2(\chi)} }+ \sqrt{1-{4\lambda'} (\chi-1){\mathbb{K}^2(\chi)}
}}\,.
\ee
In the limit of a long string, for $(-\chi)\to\infty$, the expression for
$\gamma(\lambda)$ can be expanded in powers of $4\lambda' (-\chi)
\mathbb{K}^2(\chi)\sim \lambda'\ln^2(-\chi)$. Examining the expression for
$\Delta\gamma$ one finds that its contribution to $\gamma(\lambda)$ is suppressed
by the factor $[\mathbb{E}(\chi)-\mathbb{K}(\chi)]/(-\chi\mathbb{K}(\chi))\sim
1/\ln(-\chi)$ compared to the first term in the square brackets in
\re{gamma-sqrt}. This implies that in the limit $(-\chi)\to\infty$ with
$\lambda'\ln^2(-\chi)= {\rm fixed}$ the leading asymptotic behavior of
\re{gamma-sqrt} reads
\be\label{for}
\gamma(\lambda) = L\left[\sqrt{1+{\lambda'} \ln^2(-\chi)}-1\right]+\ldots\,.
\ee
This relation resums all double-logarithmic corrections $\sim L[{\lambda'}
\ln^2(-\chi)]^n$ to the anomalous dimension $\gamma(\lambda)$ to all orders $n$
at strong coupling. In particular, for $n=1$ and $n=2$ it reproduces
\re{AsyPertExp}. For ${\lambda'} \ln^2(-\chi)\ll 1$ one expands the square-root
in \re{for} and arrives at a BMN-like expansion \re{AnomalousDimensionPertExp}.
At the same time, for ${\lambda'} \ln^2(-\chi)\gg 1$ the relation \re{for} leads
to the expression
\be
\gamma(\lambda) = L \sqrt{\lambda'} \ln (-\chi) + \ldots =
\frac{\sqrt{\lambda}}{\pi} \ln(-\chi) + \ldots \,.
\ee
For $(-\chi)\to\infty$, the dependence of $\chi$ on the ratio $N/L$ and the BMN
coupling $\lambda'$ follows from the second relation in \re{FrolovTseytlin}
\be\label{xi-eq}
\frac{1}{4\chi^2} =\frac{L^2}{N^2} \left[ \frac{1}{\ln^2 (- \chi)} +
\lambda^\prime \right]\, .
\ee
For $\xi_{\rm srt}< 1$ and $N\gg L$ one finds from \re{xi-eq} and \re{xi-str}
that $\chi\sim\chi_0 = - \frac{N}{2 L} \ln \frac{N}{L}$ and, therefore,
\be\label{gamma-str-dlog}
\gamma(\lambda)= L\left[\sqrt{1+\frac{\lambda}{(\pi L)^2} \ln^2(N/L)}-1\right] +
\ldots
\ee
For $\xi_{\rm srt}\gg 1$ one finds from \re{xi-eq} and \re{xi-str} that
$(-\chi)\sim N/(2L\sqrt{\lambda'})\sim N/\sqrt{\lambda}$ and, therefore,
\be\label{gamma-str-log}
\gamma(\lambda)=\frac{\sqrt{\lambda}}{\pi} \ln(N/\sqrt{\lambda}) + \ldots
\ee
This relation reproduces the correct asymptotic behavior of the anomalous
dimension in the regime \re{3rd} and it is in agreement with the results
of Refs.~\cite{GubKlePol03,FroTse03,Kru05}.

The coefficient in front of $\ln N$ in \re{gamma-str-log} determines the leading
asymptotics of the cusp anomalous dimension in the strong coupling regime in the
$\mathcal{N}=4$ SYM theory, Eqs.~\re{gamma=cusp} and \re{cusp}. It follows from
the relations \re{gamma-str-dlog} and \re{gamma-str-log} that quite remarkably
the cusp anomaly at strong coupling can be obtained by resumming double
logarithmic terms $\sim L [\lambda\ln^2(N/L)/L^2]^k$ in the expansion of
anomalous dimensions of operators of higher twist $L$ and large Lorentz spin
$N\gg L$.

\subsection{Two-cut solution in string sigma model}
\label{StringTwoCut}

In the previous section, we observed that for $N\gg L$ the dependence of the
anomalous dimension $\gamma (\lambda)$ on the coupling constant is different for
$\xi_{\rm str}< 1$ and $\xi_{\rm str}\gg 1$. In the former case, $\gamma
(\lambda)$ has a BMN-like expansion in powers of $\lambda/L^2$, while in the
latter case $\gamma(\lambda)$ is not analytical in $\lambda$ and scales
logarithmically $\sim \sqrt{\lambda}\ln N$. The reason for this non-analyticity
is that for $\xi_{\rm str}\gg 1$ the end-points of the rotating string approach
the boundary of the AdS space and the dominant contribution to the energy of the
string, or equivalently the anomalous dimension \re{E-gamma}, comes from the
vicinity of two spikes, $\gamma(\lambda) \sim (\sqrt{\lambda}/\pi) \cdot 2
\rho_0$. Here the radial coordinate of the spikes scales for $\xi_{\rm str}\gg 1$
as $\rho_0\sim \ft12\ln (-\chi)\sim \ft12\ln(N/\sqrt{\lambda})$, Eqs.~\re{rho0}
and \re{xi-eq}. The phenomenon is rather general~\cite{BelGorKor03,Kru05} and it
holds true for classical string configurations with an arbitrary number of spikes
$n$. In that case, each spike provides a logarithmic contribution to the energy
and the anomalous dimension scales for $\xi_{\rm str}\gg 1$ as $\gamma(\lambda)=
{n\sqrt{\lambda}}/{(2\pi)} \ln(N/\sqrt{\lambda})$. At first glance, this
mechanism is rather different from the one in gauge theory. We recall that in
gauge theory, to one-loop order, the logarithmic scaling of anomalous dimension
for $N\gg L$ arises due to collision of cuts for the spectral curve of the spin
chain, Eq.~\re{curve}. It is known that the classical equations of motion for the
string on the AdS${}_5\times$S${}^5$ background are completely
integrable~\cite{ZakMik78,ManSurWad02} and their solutions are parameterized by
the spectral curves. Moreover, for the strings on the AdS${}_3\times$S${}^1$ part
of the target space the spectral curve can be identified as a complex
hyperelliptic curve~\cite{Kri94}. For the folded rotating closed string discussed
in the previous section, it is given by the elliptic curve with symmetric
branching points on the real axis~\cite{KazZar04}
\be\label{curve-str}
\Gamma_{\rm str}: \qquad y^2=(x^2-a_{\rm str}^2)(x^2-b_{\rm str}^2)\,,
\ee
with $a_{\rm str}$ and $b_{\rm str}$ taking positive values, $b_{\rm str}< a_{\rm
str}$. In this section, we shall translate different asymptotic behavior of the
anomalous dimensions for $N\gg L$, Eqs.~\re{gamma-str-dlog} and
\re{gamma-str-log}, into properties of the curve \re{curve-str} and reveal the
mechanism responsible for the logarithmic scaling \re{gamma-str-log}.

Similar to the classical $SL(2)$ spin chain, the classical string equations of
motion admit the Lax representation and they can be solved exactly by
constructing the Baker-Akhiezer function~\cite{Kri94,ZakMik78}. As before, the
Bloch-Floquet multiplier for this function gives rise to the (quasi)momentum
$p(x)$ which is the generating function for the conserved charge including the
energy. For the folded rotating string configuration, $p'(x)$ is an analytical
function in the complex plane with two symmetric cuts $[-a_{\rm str},-b_{\rm
str}]\cup[b_{\rm str}, a_{\rm str}]$. It is uniquely fixed by the requirement
that $dp=p'(x)dx$ should be a meromorphic differential on the complex curve
\re{curve-str} satisfying the following conditions on the upper sheet of
$\Gamma_{\rm str}$~\cite{KazZar04}:
\begin{itemize}

\item Prescribed asymptotics at infinity and at the origin
\be\label{p-as}
dp \stackrel{x\to\infty}{\sim} -2\frac{dx}{x^2} \frac{E+N}{L}\,,\qquad dp
\stackrel{x\to 0}{\sim} -dx\,\frac{2}{\lambda'} \frac{E-N}{L}
\ee

\item Single-valuedness condition
\be\label{p-periodsSigma}
\int_{b_{\rm str}}^{a_{\rm str}} dp = 0\,,\qquad \int_{a_{\rm str}}^{\infty} dp =
-\pi m\,,
\ee

\item Double poles at $x=\pm\sqrt{\lambda}/(\pi L)\equiv \sqrt{\lambda'}$
\be\label{poles}
dp \sim dx
\left[
- \frac{1}{(x\pm \sqrt{\lambda'})^2}
+
\mathcal{O} ((x\pm \sqrt{\lambda'})^0) \right]
\, .
\ee

\end{itemize}
The resulting expression for the differential $dp$ takes the form
\be\label{p-ans}
dp = \frac{dx}{y}
\left[
\frac{y_+}{(x-\sqrt{\lambda'})^2}+\frac{y_+'}{x-\sqrt{\lambda'}}
+
\frac{y_+}{(x+\sqrt{\lambda'})^2}-\frac{y_+'}{x+\sqrt{\lambda'}}
+
C
\right]\,,
\ee
where $y=y(x)$ is defined in \re{curve-str}, $y_+ = y(\sqrt{\lambda'})$ and
$y_+'=y'(\sqrt{\lambda'})$.

Equation \re{p-ans} depends on three parameters ${a_{\rm str}}$, ${b_{\rm str}}$ and
$C$. They are fixed by the normalization conditions \re{p-as} and \re{p-periodsSigma}
as
\ba\label{b-str}
b_{\rm str}
\!\!\!&=&\!\!\!
\frac{1}{m\mathbb{K}(\tau)} \left[
\left(1-\frac{\lambda^\prime}{a_{\rm str}^2}\right)
\left(1 - \frac{\lambda^\prime}{b_{\rm str}^2}\right)
\right]^{-1/2}
\\[3mm]
C
\!\!\!&=&\!\!\!
-\frac{m a_{\rm str}}{2}
\left[
\mathbb{E}(\tau) -\frac{\lambda^\prime}{a_{\rm str}^2} \mathbb{K}(\tau) \right]
\, , \nonumber
\ea
with the modular parameter $\tau = 1-{b_{\rm str}^2}/{a_{\rm str}^2}$. In
addition, one finds from \re{p-as} the following expressions for the ratio
$N/L$ and for the anomalous dimension $\gamma(\lambda)=E-N-L$
\ba\label{N/L}
{N}/{L} \!\!\!&=&\!\!\!
\frac{m}{2}
\left[
\mathbb{E}(\tau)\lr{ a_{\rm str} + \frac{{\lambda'}}{b_{\rm str}} }
-
\mathbb{K}(\tau) \lr{ b_{\rm str} + \frac{{\lambda'}}{a_{\rm str}} }
\right]
\, , \\[3mm]
\gamma(\lambda)/L \!\!\!&=&\!\!\!
m \left[ \mathbb{K}(\tau)\, b_{\rm str} - \mathbb{E}(\tau) \frac{{\lambda'}}{b_{\rm str}} \right]
- 1
\, , \nonumber
\ea
with $b_{\rm str}$ defined in \re{b-str} and $a_{\rm str}=b_{\rm str}/\sqrt{1-\tau}$.

Let us examine the dependence of the anomalous dimension on the coupling constant
$\lambda'=\lambda/(\pi L)^2$. Assuming that $b_{\rm str}$ and $\tau$ both admit a
regular expansion in powers of $\lambda'$, one substitutes into \re{b-str} and
\re{N/L}
\be\label{b-exp}
b_{\rm str} = b_{\rm str}^{(0)} + \lambda^\prime \, b_{\rm str}^{(1)} +  + \ldots
\, , \qquad \tau = \tau^{(0)} + \lambda^\prime \,\tau^{(1)} + \ldots \, ,
\ee
and matches the coefficients in front of powers of $\lambda'$. In this way, one
obtains to leading order
\be\label{b0}
b_{\rm str}^{(0)} = \frac{1}{m\mathbb{K}(\tau^{(0)})}
\,,\qquad
\frac{N}{L} =
\frac12\left[\frac{\mathbb{E}(\tau^{(0)})}{\sqrt{1-\tau^{(0)}}\mathbb{K}(\tau^{(0)})}
-1\right]\,,
\ee
and all subleading corrections to \re{b-exp} are expressed in terms of the leading
terms. Then, the first few coefficients of the BMN series for anomalous dimension
\re{AnomalousDimensionPertExp} are given by
\ba
\label{gamma1Diff} \gamma^{(0)} \!\!\!&=&\!\!\! \frac{m^2}2 \mathbb{K}
(\tau^{(0)}) \left[ (2 - \tau^{(0)}) \mathbb{K} (\tau^{(0)}) - 2 \mathbb{E}
(\tau^{(0)}) \right] \, ,
\\
\label{gamma2Diff} \gamma^{(1)} \!\!\!&=&\!\!\! \frac{m^4}{8} \mathbb{K}^{3}
(\tau^{(0)}) \left[ \lr{ 4 (2 - \tau^{(0)}) \sqrt{1 - \tau^{(0)}} - (\tau^{(0)})^2}
\mathbb{K} (\tau^{(0)}) - 8 \sqrt{1 - \tau^{(0)}} \mathbb{E} (\tau^{(0)}) \right]
\, .
\ea
Together with the second relation in \re{b0}, they define the parametric dependence
of the anomalous dimension \re{AnomalousDimensionPertExp} on $N/L$.

Equations \re{b0} determine perturbative corrections to the anomalous dimension of
long scalar operators in the $\mathcal{N}=4$ SYM theory. According to \re{match},
it should match at one-loop order a similar asymptotic expression \re{Energy-2-cut}
obtained on the gauge theory side within the semiclassical approach. The conformal
spin of scalars equals $s=1/2$ and the scaling parameter $\beta=L/(L+2N)$, Eq.\
\re{beta}, is given by $\beta = \sqrt{1 -\tau^{(0)}}\,{\mathbb{K}(\tau^{(0)})}/
{\mathbb{E}(\tau^{(0)})}$ in agreement with \re{ab}. Then, one observes that
$\gamma^{(0)}$ and $\varepsilon$ given by \re{gamma1Diff} and \re{Energy-2-cut},
respectively, verify the relation \re{match}. The expressions for the branching
points $b_{\rm str}^{(0)}$ and $b$, defined in \re{b0} and \re{ab}, respectively,
are different, $b/b_{\rm str}^{(0)} = \beta/2$, but the agreement can be restored
through the rescaling of the local complex parameter $x$ in the definition of the
curve \re{curve-str}, $x \to x \beta/2$.

The functional form of the anomalous dimensions \re{N/L} is different compared to
the ones found in the previous section, Eq.~\re{FrolovTseytlin}. The agreement is
achieved by means of the Landen transformation of the modular parameters
\cite{BeiFroStaTse03}
\be
\label{LandenGauss}
\chi^{(0)} = - \frac{\left( 1 - \sqrt{1 - \tau^{(0)}} \right)^2}{4 \sqrt{1 -
\tau^{(0)}}} \, ,
\ee
upon which the relations \re{N/L} and \re{FrolovTseytlin} coincide provided that
$m=1$. Remember that $(-\chi)$ depends on the radial coordinate of the spike
$\rho_0$, Eq.~\re{rho0}, so that $\sqrt{1-\tau}=\e^{-2\rho_0}$ and the  long
string limit $\rho_0\to\infty$ corresponds to $\tau\to 1$. We have demonstrated
in Sect.~\ref{CollisionCuts} that on the gauge theory side the limit $\tau\to 1$
corresponds to $a\to 1/(2m)$ and $b\to 0$, Eq.~\re{ab-large}. As a consequence,
the two cuts $[-a,-b]$ and $[b,a]$ collide at the origin yielding the logarithmic
scaling of the one-loop anomalous dimension \re{E-m}. Let us examine the limit
$\tau\to 1$ of the obtained stringy expressions \re{b-str} and \re{N/L}.

Since $\mathbb{K}(\tau)\sim -\ft12 \ln(1-\tau)$ for $\tau\to 1$, one would expect
from \re{b-str} that $b_{\rm str}$ should vanish in this limit. Indeed, this is
the case for $\lambda'=0$ while for $\lambda'\neq 0$ one deduces from \re{b-str}
that the reality condition for $b_{\rm str}$ implies that its minimal value is
bounded as $b_{\rm str}\ge \sqrt{\lambda'}$. Carefully examining \re{b-str} for
$\tau\to 1$ one finds
\be\label{barrier}
b_{\rm str}=\sqrt{{\lambda'}+\frac{1}{m^2\ln^2 \sqrt{1-\tau}}}\,.
\ee
For $\lambda'=0$ this relation coincides with the one-loop expression
$b^{(0)}_{\rm str}=2 b/\beta$, Eq.~\re{ab-large}. Matching \re{barrier} into
\re{b-exp} we conclude that higher order corrections to $b_{\rm str}$ push its
minimal possible value away from the origin and, therefore, prevent the two cuts
$[-a_{\rm str},-b_{\rm str}]$ and $[b_{\rm str},a_{\rm str}]$ to collide. From
\re{N/L} one finds in the limit $\tau\to 1$
\be\label{gamma-str}
\beta\approx\frac{L}{2N}
=
\frac{\sqrt{1-\tau}}{m\, b_{\rm str}}
+ \ldots
\,,\qquad
\gamma(\lambda) = L \left[- m \, b_{\rm str} \, \ln{\sqrt{1-\tau}} - 1\right]
+
\ldots
\, .
\ee
The asymptotic behavior of these expressions depends on the value of the
parameter $\xi_{\rm str}=\lambda'\ln^2(N/L)$, Eq.~\re{xi-str}.

For $\xi_{\rm str}< 1$ the expression for $b_{\rm str}$, Eq.~\re{barrier}, admits
a series expansion in $\lambda'$ and leads together with \re{gamma-str} to
\be\label{ab-str}
b_{\rm str}=\frac{1}{m\ln\frac{N}{L}} + \ldots \,,\qquad a_{\rm
str}=\frac{N}{2m{L}} + \ldots \,,\qquad \sqrt{1-\tau}=\frac{L/N}{2 \ln
\frac{N}{L}} + \ldots \,.
\ee
Substitution of \re{barrier} into \re{gamma-str} yields an expression for the
anomalous dimension $\gamma(\lambda)$ which coincides with \re{gamma-str-dlog}
for $m=1$. It is instructive to compare the positions of the cuts in gauge
theory, $a$ and $b$, and on the string side, $\widehat a=\beta a_{\rm str}/2$ and
$\widehat b=\beta b_{\rm str}/2$. Here the additional factor $\beta/2$ appears
due to a different definition of the local parameter $x$ on the spectral curves
\re{curve-reduced} and \re{curve-str}. In this way we find $ \widehat a
=1/{(2m)}+\ldots$ and $\widehat b={(L/N)}/{(4m\ln(N/L))}+\ldots$ which coincides
with the similar relation \re{ab-large} in gauge theory to one-loop order.

For $\xi_{\rm str}\gg 1$ the expression for $b_{\rm str}$, Eq.~\re{barrier}, is
not analytical in the BMN coupling $\lambda'=\lambda/(\pi L)^2$
\be
b_{\rm str}=\sqrt{\lambda'} + \ldots
\,,\qquad
 a_{\rm
str}=\frac{N}{2m{L}} + \ldots \,,\qquad \sqrt{1-\tau}=\frac{Lm}{2N}
\sqrt{\lambda'} + \ldots \,,
\ee
while the expression for $a_{\rm str}$ is the same as in \re{ab-str}. This
suggests that for $\xi_{\rm str}\gg 1$ higher order corrections only modify the
lower edge of the cut. One finds from \re{gamma-str} that the anomalous dimension
scales as $\gamma(\lambda)\sim m\sqrt{\lambda}\ln (N/(m\sqrt{\lambda}))$ and
matches \re{gamma-str-log} for $m=1$. We conclude that the logarithmic scaling of
the anomalous dimension for $N\gg L$ on the string side is realized when the
inner boundary of the cut $b_{\rm str}$ approaches its minimal possible value
$\sqrt{\lambda'}$  (see Fig.~\ref{CollisionCutsFig}c) which coincides with the
position of the double pole of the differential \re{poles}.

\section{Conclusion}
\label{ConclusionSection}

In the present paper we have studied the properties of anomalous dimensions
of Wilson operators of higher twist $L$ and large Lorentz spin $N$ in the
weak and strong coupling regimes by making use of the remarkable integrability
symmetry on both sides of the gauge/string correspondence. We concentrated on
operators which have the minimal anomalous dimension for given Lorentz spin
and put a special emphasis on the appearance of the single-logarithmic behavior,
$\gamma(\lambda) \sim \ln N$, in the thermodynamic limit $L\to\infty$.

On the gauge theory side, we applied the method of the Baxter $Q-$operator to
identify different regimes of the minimal anomalous dimension in integrable
sectors of (supersymmetric) Yang-Mills theory to one-loop order. We argued that
for $N \gg L$ the asymptotic behavior of $\gamma(\lambda)$ is controlled by the
parameter $\xi=\ln(N/L)/L$. For $\xi < 1$ the anomalous dimension possesses the
BMN scaling $\gamma(\lambda)\sim \lambda/L$, while for $\xi \gg 1$ it scales
logarithmically with $N$. Transition to the second, logarithmic regime manifests
itself through the divergence of the semiclassical expansion for $\gamma(\lambda)$
as $\xi \sim 1$. The anomalous dimension is uniquely determined by the configuration
of Bethe roots which condense in the thermodynamic limit on two symmetric cuts. We
demonstrated that the semiclassical approach breaks down for $\xi\sim 1$ due to
the collision of cuts at the origin and worked out an asymptotic expression for
anomalous dimensions which is valid throughout the entire region of $\xi$.

On the string theory side, we used the identification of the minimal anomalous
dimension of scalar operators in the $\mathcal{N}=4$ SYM at strong coupling as
the energy of folded string rotating on AdS${}_3\times$S${}^1$ part of the target
space. Similar to the previous case, the anomalous dimension has different
behavior for $N\gg L$ depending on the value of the parameter $\xi_{\rm str} =
\lambda \ln^2(N/L)/L^2$. For $\xi_{\rm str}<1$ the anomalous dimension has a
regular expansion in powers of the BMN coupling and its lowest term matches the
one-loop expression for $\gamma(\lambda)$ at weak coupling for $\xi<1$. For
$\xi_{\rm str}\gg 1$ the anomalous dimension scales logarithmically but its
dependence on the 't Hooft coupling is not analytical anymore. We described the
latter regime using two different (although equivalent) configurations. In terms
of the folded rotating string the logarithmic scaling occurs when two most
distant points of the string (two spikes) approach the boundary of the AdS space.
In terms of the spectral curve for the classical string sigma model, the same
configuration is described by the elliptic curve with symmetric branching points.
Different regimes of $\gamma(\lambda)$ arise depending on the position of the
branching points. In the logarithmic regime, the inner branching points approach
the minimal possible value $\pm \sqrt{\lambda^\prime}$ so that the anomalous
dimension ceases to obey the BMN scaling.

Integrability played a key role in our analysis. In generic (supersymmetric)
Yang-Mills theory it holds to one-loop order for Wilson operators belonging to
special, holomorphic sectors only. We would like to stress that logarithmic
behavior of anomalous dimensions is not tied to integrability. In non-integrable
sectors the mixing matrix for Wilson operators contains additional terms which
break integrability symmetry. They do not affect however the logarithmic scaling
of anomalous dimension for $N\to\infty$. The reason for this is that the
logarithmic scaling of anomalous dimension can be associated with the
contribution of soft gluons (i.e., gauge field quanta). Soft gluon radiation is
not sensitive to the quantum numbers of Wilson operators (except of the total
Lorentz spin) and, therefore, it provides the same logarithmic contribution to
the anomalous dimension of Wilson operators in all sectors. Integrability allows
one to identify various regimes of the asymptotic behavior of anomalous
dimensions and to determine the ``critical'' value of $\ln(N/L) \sim L $ at which
the logarithmic scaling sets in.

As a function of the total Lorentz spin $N$, the anomalous dimensions of
twist$-L$ operators occupy a band. Our discussion was restricted to the minimal
anomalous dimensions belonging to the lower edge of the band. It would be
interesting to extend the above analysis to excited states
\re{SingleLogAsymptotics} with $m > 1$ and describe the band structure occupied
by the anomalous dimensions as it arises form the Baxter equation on the gauge
theory side. In string theory, there are two different classical configurations
yielding the same logarithmic asymptotics of anomalous dimensions at strong
coupling --- the multiple folded string \cite{FroTse03} and the spiky string
\cite{Kru05}. The coefficient in front of the logarithm, $m$, is twice the number
of foldings of the string on itself, in the former case and it is the number of
spikes, in the latter one. It would be interesting to construct a generic
configuration which interpolates between both solutions.

\vspace{0.5cm}

\noindent {\bf Note added}: After the work has been completed we were informed by
Yuji Satoh that he, in collaboration with Kazuhiro Sakai, analyzed the spectrum
of anomalous dimensions with a special emphasis on the appearance of logarithmic
scaling in the large spin limit. For the two-cut solution, their findings are in
agreement with the analysis presented in sections \ref{WKBloSection} and
\ref{StringTwoCut} of this paper. In particular, they also came to the conclusion
that, in the thermodynamical limit $L\to\infty$, the semiclassical approach to
the Bethe Ansatz equations is only applicable for $\ln (N/L) < L$ and one can not
reproduce $\sim \ln N$ behavior of anomalous dimensions unless the finite size
corrections in $1/L$ are included.

\vspace{0.5cm}

We would like to thank to thank S.~Derkachov, Yu.~Makeenko, A.~Manashov, Y.~Satoh,
F.~Smirnov, E.~Sokatchev, B.~Stefanski and A.~Tseytlin for very useful discussions
and correspondence. A.G.\ is grateful to Laboratoire de Physique Th\'eorique (Orsay)
for hospitality extended to him during his stay which was partially supported by
the Russian-French Exchange Program. This work was supported by the U.S.\ National
Science Foundation under grant no.\ PHY-0456520 (A.B.), by the grant CRDF-RUP2-261-MO-04
from the U.S.\ Civilian Research and Development Foundation and Russian Foundation
for Basic Research under contract RFBR-04-011-00646 (A.G.).

\vspace{0.5cm}

\appendix
\setcounter{section}{0}
\setcounter{equation}{0}
\renewcommand{\theequation}{A.\arabic{equation}}

\section*{Appendix \ \   Asymptotic expression for the energy}

The asymptotic expression for the energy \re{energy-as} involves roots of the
transfer matrix and is not well suited for performing the thermodynamical limit
$L \to \infty$. Let us rewrite the energy in terms of the transfer matrix itself.
We notice from \re{t-roots} that
\be
\frac{d}{du} \ln t_L(u) = \sum_{k=1}^L \frac1{u-\delta_k}
\ee
and replace the $\psi-$functions in \re{energy-as} by series representation
\be
\psi(s+i\delta_k) + \psi(s-i\delta_k) -2 \psi(2s) = \sum_{k=0}^\infty
\left[\frac{i}{i(s+k)-\delta_k}+\frac{-i}{-i(s+k)-\delta_k} -\frac{2}{2s+k}
\right]\,.
\ee
Then, the energy \re{energy-as} can be written as an infinite sum
\be\label{E-main}
E = 2\ln 2-\sum_{k=0}^\infty \frac{d}{dk}f(k)\,,\qquad
f(x)=\ln \left[ \frac{t_L(i(s+x))t_L(-i(s+x))}{(2s+x)^{2L}} \right]\,,
\ee
where $f(x)\to 2\ln 2$ for $x\to\infty$. The sum can be evaluated with a help of
the Euler-Maclaurin summation formula
\be\label{E-ex}
E = f(0)-\frac1{2} f'(0)+\sum_{k=1}^\infty \frac{B_{2k}}{(2k)!}f^{(2k)}(0) =
\frac{\textrm{d}}{\e^\textrm{d}-1} f(0)
\ee
with $B_{2k}$ being Bernoulli numbers. Taking $u=\pm i(s+x)$ in the Baxter
equation \re{Baxter-eq}, one obtains $t_L(\pm i(s+x))$ as the ratio of
$Q-$functions and finds for $x\to 0$
\be
f(x)=\ln \left[\frac{Q(i(s+x+1))}{Q(i(s+x))}
\frac{Q(-i(s+x+1))}{Q(-i(s+x))}\right] + \mathcal{O}\lr{ x^{L}}\,.
\ee
Then, one uses the WKB ansatz $Q(u) \sim \exp(\eta^{-1}S(u\eta))$ to get
\be
f(x)= \frac1{\eta}\big[S\lr{i{(s+x+1)}{\eta}}-S\lr{i{(s+x)}{\eta}}
+S\lr{-i{(s+x+1)}{\eta}}-S\lr{-i{(s+x)}{\eta}} \big]+ \mathcal{O}\lr{ x^{L}}
\, .
\ee
Substituting this expression into \re{E-ex} one finally obtains the relation
\be\label{E-app}
E=f(0)-\frac1{2} f'(0)+ \ldots = -{2s}{\eta} S''(0)+\ldots= -\frac{2\beta}{L}
S''(0)+\ldots\,,
\ee
which coincides with the semiclassical expression \re{E-naive-expansion}. It is
important to keep in mind that Eq.~\re{E-app} was obtained under the assumption
that contribution of terms with higher derivatives is small, $f''(0) \ll f'(0)$,
or equivalently $ \eta S'''(0)\ll S''(0)$. This relation is not satisfied if the
distribution of Bethe roots scales at the origin as $\sim \ln x$,
Eq.~\re{S0-log}. In that case, one has to rely on the formula \re{E-main} which
resums all singular terms.



\end{document}